\documentclass[aps,prd,eqsecnum,11pt,showpacs,preprintnumbers,nofootinbib]{revtex4}
\usepackage{graphicx}
\usepackage{caption}
\usepackage{multirow}
\usepackage{float}
\usepackage{subfig}
\usepackage{amssymb,amsmath}
\newcommand{\bes}{\begin{subequations}}
\newcommand{\ees}{\end{subequations}}
\newcommand{\be}{\begin{equation}}
\newcommand{\ee}{\end{equation}}
\newcommand{\bea}{\begin{eqnarray}}
\newcommand{\eea}{\end{eqnarray}}

\newcommand{\la}{\langle}
\newcommand{\ra}{\rangle}
\def\frc#1#2{{\textstyle{\frac{#1}{#2}}}}
\begin{document}

\title{Future singularities if the universe underwent Starobinsky inflation in the past}
\author{Eric D.~Carlson$^{\rm a}$}
\altaffiliation{\tt ecarlson@wfu.edu}
\author{Paul R.~Anderson$^{\rm a}$}
\altaffiliation{\tt anderson@wfu.edu}
\author{John~R.~Einhorn$^{\rm a}$}
\author{Bradley~Hicks$^{\rm a}$}
\author{Andrew~J.~Lundeen$^{\rm a}$}
\affiliation{${}^{a)}$Department of Physics, Wake Forest University,
Winston-Salem, North Carolina 27109, USA,}

\begin{abstract}

The effects which quantum fields and an $\alpha_0 R^2$ term in the gravitational Lagrangian have on
future singularities are investigated.  While all values of $\alpha_0$ are considered, an emphasis is placed on those values which are compatible with the
 universe having undergone Starobinsky inflation in the past.  These are also values which lead to stable solutions to the semiclassical backreaction equations in the present universe.  The dark energy is modeled as a perfect fluid, and the focus is on type I-IV singularities and little rips which result
when the classical Einstein equations are solved with various types of dark energy as a source.
First, evidence is provided that the energy densities of massive conformally coupled scalar fields approach that of the conformally invariant scalar field as a type III singularity is approached.  Then, solutions to the semiclassical backreaction equations are investigated when conformally invariant fields and the $ \alpha_0 R^2$ term in the gravitational Lagrangian are present.  General proofs regarding the behaviors of the solutions are given. The proofs are illustrated by analytic and numerical calculations in specific cases.

\end{abstract}


\maketitle

\newpage

\section{Introduction}
\label{introduction}

The 2015 analysis by the Planck Collaboration has provided significant constraints on many inflationary scenarios~\cite{planck-2015-inflation}.
One scenario that is still favored by the data is Starobinsky inflation, in which inflation is driven by an $R^2$ term in the
gravitational Lagrangian~\cite{starobinsky-1,starobinsky-2}.  The analysis also provided~\cite{planck-2015-w} a revised constraint on the dark energy equation of state parameter $w = p_{de}/\rho_{de}$, namely $w = -1.006 \pm 0.045 $.  This is, of course, consistent with a cosmological constant ($w = -1$).  But it weakly favors phantom dark energy ( $w < -1$),
which can lead to a future singularity.  For example, a big rip singularity occurs if $w < -1$ is a constant~\cite{big-rip}.

Predictions of the existence of final singularities are usually made based on solutions to the classical Einstein equations with various dark matter sources for which $w < -1$.
The existence of singularities usually indicates the breakdown of classical general relativity, which must be replaced by a quantum theory of gravity such
as string theory or loop quantum gravity.  While the full quantum theory of gravity is as yet unknown, there is a semiclassical approximation which should be valid, at least in many cases, when spacetime curvatures are smaller than the Planck scale.  In semiclassical gravity, renormalization inevitably predicts the existence of an $\alpha_0 R^2$ term in the gravitational Lagrangian~\cite{b-d-book}.

In this paper, we undertake an investigation of quantum effects on future singularities of types I-IV and on little rip models, all of which are defined below.
The dark energy which causes these singularities is assumed to be a perfect fluid with equation of state $p_{de} = p_{de}(\rho_{de})$.
We consider the effects of conformally invariant free quantum fields in the conformal vacuum state, and in one case we consider the effects of free massive conformally coupled scalar fields.
We restrict our attention to semiclassical gravity along with an $\alpha_0 R^2$ term in the gravitational Lagrangian  because it is both the most conservative way and the most developed way to take quantum effects into account.
Some of the terms in the stress-energy tensors for the quantum fields have the same form as those from the $\alpha_0 R^2$
term.  This results in an effective coefficient for the $R^2$ term which we call $\alpha$.  Details are given in Sec.~II. We consider all values of $\alpha$ but emphasize the values which are compatible with Starobinsky inflation.

One serious limitation of semiclassical gravity is that the solutions are not expected to be reliable once the spacetime curvature reaches the Planck scale.
However, if Starobinsky inflation occurred, then $\alpha \sim 10^9$~\cite{huang}, which is large enough that quantum effects can be important when the spacetime curvature is well
below the Planck scale.  As we shall show, even if quantum effects are important that does not guarantee that a final singularity will be removed by backreaction effects.
Nevertheless it can still be of some interest to investigate the predictions of the semiclassical approximation
 regarding the final singularity.  This is exactly what is done when studying singularities in classical general relativity.  In the semiclassical case, such an approach was taken in~\cite{ashtekar} for Callan-Giddings-Harvey-Strominger, CGHS, black holes
which form from collapse in (1+1) dimensions~\cite{CGHS}. In cases where the singularities are not removed, we study the singularity structure of the solutions to the semiclassical backreaction equations to
determine if the classical singularity has been altered, and whether it has been weakened or strengthened.

\subsection{Types of final singularities}

Previously, several types of possible final singularities have been identified and studied.  Big rip singularities, in which the scale factor and its derivatives (and hence the energy density and pressure) approach infinity in a finite proper time, were discussed in~\cite{big-rip}.  As the singularity is approached, tidal forces become arbitrarily large, and all bound objects are torn apart.
Sudden singularities, in which the scale factor and its first derivative (and hence the energy density) are finite and higher derivatives diverge in a finite proper time, were discussed in~\cite{b-g-t,barrow}, with examples relevant to inflation given in~\cite{barrow-graham}. Examples of other types of future singularities are found in~\cite{barrow-graham,barrow-2,nojiri-odintsov,barrow-3,stefancic}.  One can also consider cases such as the little rip~\cite{little-rip}, in which there is no singularity because the scale factor and its derivatives approach infinity only at an infinite proper time in the future.  However, at finite times, tidal forces become arbitrarily large, and bound objects are torn apart.

The classification of final singularities that we use is given in~\cite{not}.
For type I singularities, also known as big rip singularities, the scale factor and the spacetime curvature diverge at a finite proper time.  For types II-IV the scale factor is finite
at the singularity, which occurs at a finite proper time.  For type II singularities, also known as sudden singularities, the energy density and the first derivative of the scale factor are finite, but the pressure and the second derivative of the scale factor diverge.  For type III singularities the first derivative of the scale factor, the energy density, and the pressure diverge.  Type IV singularities are the weakest, with only the derivative of the pressure and the third derivative of the scale factor diverging.  Even milder singularities in which the lowest order derivative of the scale factor that diverges is four or larger can also be considered, and are sometimes categorized as type IV.  However, we will restrict the definition of type IV to include only singularities in which the first two derivatives of the scale factor are finite and the rest diverge.  Other classification schemes have also been discussed~\cite{class-1,class-2,class-3,class-4}.

\subsection{Previous work relating to quantum effects on final singularities}

Work has been done regarding the effects of quantum cosmology on final singularities~\cite{bouhmadi-moniz,sami-singh-tsujikawa,barboza-lemos,dabrowski-kiefer-sandhofer,
kamenshchik-kiefer-sandhofer,bouhmadi-keifer-sandhofer-moniz,kiefer-1,kiefer-2,kamenshchik-manti,deharo,bamba-deharo-odintsov,kamenshchik,bouhmadi-keifer-kraemer,
albarran-bouhmadi,albarran-bouhmadi-cabral-martin} as well as the effects of $F(R)$, $F(G)$, and similar theories~\cite{abdalla-nojiri-odinstov,srivastava,
nojiri-odinstov-1,bamba-nojiri-odintsov,capozziello-laurentis-nojiri-odintsov,bamba-odintsov-sebastiani-zerbini,nojiri-odintstov-2,appleby-battye-starobinsky,
lopez-elizalde,nojiri-odintsov-3}.  However, as mentioned above, our focus in this paper is on quantum effects on final singularities due to semiclassical gravity.
The question of whether quantum effects due to semiclassical gravity can remove these singularities has been addressed in two ways.  The first and easiest way is to
compute $\langle T_{ab} \rangle$ for various quantum fields in a given background spacetime with a final singularity and compare it with the stress energy of the dark energy which is responsible for the singularity.  One expects semiclassical backreaction effects to be unimportant if the stress energy of the dark energy is much larger in magnitude than that of the quantum fields.  The second way is
to actually solve the semiclassical backreaction equations when conformally invariant quantum fields are present.

Background field calculations have been done in spacetimes with type I and type II singularities for conformally invariant fields, the massless minimally coupled scalar field, and massive conformally and minimally coupled scalar fields.  For type I singularities the effects of conformally invariant scalar, spinor, and vector fields in spacetimes with constant values of $w$ were investigated in~\cite{calderon-hiscock}.  Comparing the stress energy of the quantized fields with that of the dark energy, it was found that the results vary depending on the value of $w$ and on the values of the renormalization parameters for the fields.  For values of $w$ that
are realistic for our universe, they found that quantum effects seem to strengthen the singularity.  In~\cite{ba} it was shown that the energy density
of one conformally invariant scalar field remains small compared to the dark energy up to the Planck scale for values of $w$ that are realistic for our universe.

The effects of particle production due to a massless minimally coupled scalar field on type I singularities with constant values of $w$ were investigated in~\cite{bfh} using a state for which Bunch and Davies~\cite{bunch-davies} had previously computed the stress-energy tensor.  It was found that the energy density of the created particles never dominates the dark energy density.  For this case an approximation to the full stress-energy tensor, which also includes vacuum polarization effects, was computed in~\cite{abfh}. It was found that quantum effects are important if you go close enough to the final singularity.  The same conclusion was reached in~\cite{ba} using the exact energy density which
Bunch and Davies had computed for their state~\cite{bunch-davies}.  It was also found that for a single scalar field quantum effects are not important before the Planck scale is reached.  The energy density was computed for an arbitrary fourth order adiabatic state, and it was shown that the state investigated by Bunch and Davies is an attractor in the sense that the energy density for all other fourth order adiabatic states approaches the energy density for this state as the final singularity is approached.

Both the number density of created particles and the stress-energy tensor for a conformally coupled massive scalar field were computed for the case $w = - \frc53$ in~\cite{pavlov}.  Backreaction effects were found to be unimportant for masses much smaller than the Planck mass and times which are early enough that the time until the big rip occurs is greater than the Planck time.  The energy densities of conformally and minimally coupled massive scalar fields in big rip spacetimes with constant $w$ were numerically computed for specific values of $w$ and specific states of the quantum fields in~\cite{ba}.  For conformal coupling it was found in each case considered that at late times the energy density approaches that of the conformally invariant scalar field.  For minimal coupling it was found in every case considered that at late times the energy density approaches that of the massless minimally coupled field in the state found by Bunch and Davies.

For type II singularities the effects of conformally invariant fields in certain cases using a background field approach were investigated in~\cite{calderon}.  It was found
that whether the singularity is strengthened or weakened depends on the sign of one of the renormalization parameters.
In~\cite{bbfh} the effects of particle production in models with sudden singularities when a massless,
minimally coupled scalar field is present were investigated.  It was found that particle production effects are never important near the singularity, because
the stress energy of the produced particles remains small in comparison with that of the dark energy.
The effects of particle production due to a massive conformally coupled scalar field near a sudden singularity were investigated in~\cite{bbfhd}.  Using an approximate calculation for the energy density of the produced particles, it was found that particle production effects are never important near the singularity.

To date, the only semiclassical backreaction calculations that have been done which are related to final singularities have been for conformally invariant fields.
Calculations for type I singularities have been done in~\cite{el-no-od,nojiri-odintsov-4, el-no-od-wa, nojiri-odintsov, not, srivastava, comment, hae-1, haro, hae-2}\footnote{We agree with the finding of~\cite{el-no-od-wa} that in certain cases quantum effects can eliminate the big rip singularity.  However, one of our proofs in Sec.~VII states that if the big
 rip singularity is removed by backreaction effects from conformally invariant fields and an $R^2$ term in the gravitational Lagrangian, then what happens instead is that it undergoes a bounce.  This is contrary to the suggestion in~\cite{el-no-od-wa} and also in the review article~\cite{nojiri-odintsov-3} that if the big rip singularity is removed, the de Sitter solution to the semiclassical backreaction equations replaces it.}.  They have been done for type II singularities
in~\cite{nojiri-odintsov, not,hae-3,houndjo, haro} and for type III singularities in~\cite{not,haro}.  It was found that backreaction effects can be significant, often resulting in the avoidance or softening of singularities.

\subsection{Overview of the work done in this paper}

In spite of the significant amount of work that has been done on the effects which semiclassical gravity has on final singularities, a complete and comprehensive
study of them has not yet been given.  In this paper we make significant progress toward that goal
 by investigating the behaviors of solutions
to the semiclassical backreaction equations
for conformally invariant free quantized fields and an $\alpha_0 R^2$ term in the gravitational Lagrangian when the dark energy is modeled as a perfect fluid and the spacetime is
homogeneous, isotropic, and spatially flat. We consider cases in which solutions to the classical Einstein equations with the dark energy as a source consist
of spacetimes with little rips or with future singularities of type I - IV.
In each case we develop
model-independent proofs of the behaviors of solutions to the equations.

In addition we find analytical and numerical solutions to the semiclassical backreaction equations for certain specific models of the dark energy.
These are used to illustrate the proofs and, in some cases, to compare with numerical solutions that we also obtain for
the order-reduced semiclassical equations for which the higher derivative terms have been eliminated.  The order-reduced semiclassical equations are discussed in more detail below.

We use the background field method to numerically investigate the behaviors of the energy densities of massive conformally coupled scalar fields
in a spacetime with a type III singularity.  We also use this method to analytically investigate the energy densities for conformally invariant fields in all
spacetimes with little rips or with future singularities of types I - IV.

 In many cases our results confirm those of previous authors.

\subsubsection{General analytic proofs}

The specific assumptions in our proofs regarding solutions to the semiclassical backreaction equations are as follows:  We work in spatially flat, homogeneous, and isotropic spacetimes.  We include conformally invariant quantum fields in the conformal vacuum state.  We include an $\alpha_0 R^2$ term in the gravitational Lagrangian.  The only classical matter we include is the dark energy which we assume is a perfect fluid.  We investigate all equations of state of the dark energy which result in solutions to the classical Einstein equations with future singularities of types I, II, III, and IV, or which result in spacetimes in which a little rip occurs.  For all values of the effective coefficient $\alpha$ defined in Sec.\ II, we give analytic proofs which describe the behaviors of the corresponding solutions to the semiclassical backreaction equations in all cases in which the equation of state of the dark energy results in one of these final behaviors.  In this way we are able to make definitive predictions regarding the effects of quantum fields and the $\alpha_0 R^2$ term on these singularities.

We restrict our attention to spatially flat spacetimes, because the spatial curvature will not have any significant influence on the behavior of solutions near a final singularity or at late times in a spacetime ending in a little rip.  We do not include any regular classical matter or radiation in our calculations, nor do we include a cosmological constant, because the energy density and pressure of the dark energy will dominate over all of these near a final singularity.  For conformally invariant fields in any homogeneous and isotropic state other than the conformal vacuum state there is an
extra term in the stress-energy tensor which is of the same form as that of classical radiation~\cite{wald}.  This term will be unimportant near a final singularity.

One reason we include only conformally invariant quantum fields is the evidence discussed in Sec.~\ref{sec:background-field-massive} that the energy density for the conformally coupled massive field approaches that of the massless one near the singularity.  We expect that this will be true for spin $\frc12$ and spin $1$ massive fields as well.  To see why, note that if one uses point splitting~\cite{christensen-76,christensen-78} the renormalization counterterms contain higher derivative terms which are not multiplied by any power of the mass.  Therefore they exist for both massive and massless fields.  Near the final singularity the higher derivative terms in the stress-energy tensor tend to dominate.
Most fields in the standard model of particle physics or in various grand unified theories with or without supersymmetry are either spin $\frc12$ or spin $1$.  In the massless limit when
interactions are ignored, these become conformally invariant.
We are ignoring minimally coupled fields, because it is likely there is only a small number of them compared to the large number of conformally coupled fields.  In fact, the only two likely candidates are the Higgs field, if it is a massive minimally coupled scalar field, and the graviton field.  The latter in a Robertson-Walker universe in a particular gauge can be modeled as two massless minimally coupled scalar fields~\cite{grischuck,ford-parker}.

We include an $\alpha_0 R^2$ term in the gravitational Lagrangian in part because one is required for renormalization of the quantum fields~\cite{b-d-book}. Further, as mentioned above, such a term is necessary for Starobinsky inflation to occur.  The other unique term necessary for renormalization is a Weyl squared term, but this gives a vanishing contribution to the semiclassical backreaction equations in any homogeneous and isotropic spacetime.
As discussed above, our approach is a relatively conservative one, so we do not include the effects of other higher order terms that could be present in the gravitational Lagrangian, and we restrict our attention to semiclassical effects rather than the full quantum gravity effects that occur in string theory or loop quantum gravity.

The results from our analytic proofs regarding the behaviors of solutions to the semiclassical backreaction equations are mixed.
If the universe underwent Starobinsky inflation ($\alpha > 0$) then, as was shown in~\cite{fhh}, there are no bounce solutions and thus the universe cannot
avoid a final singularity if one is present classically.  We find that if there is a classical little rip, it is turned into a big rip singularity and that a classical big rip singularity always remains a big rip singularity.  Depending on the properties of the dark energy, type III singularities either remain type III singularities or are softened to type II or type IV singularities.  Type II and IV singularities are weakened to the point that they are effectively removed, since at least the first three derivatives of the scale factor remain finite at the singularity.  However, for type II the fourth and all higher derivatives of the scale factor diverge and for type IV the fifth and all higher derivatives of the scale factor diverge.

For $\alpha < 0$, if there is classically either a big rip singularity or a little rip, then it will always be avoided by a bounce, in which the universe reaches a finite maximum size and
starts to contract.  The results for type III singularities are model dependent.  In some cases we find that the singularity is always avoided by a bounce.  In others it can be avoided by a bounce, but if this does not happen then it is weakened to a type II or type IV singularity.  If classically there is a type II or type IV singularity, then it can also be avoided by a bounce.  If this does not happen then, as for the case $\alpha >0$, quantum effects will effectively remove these singularities.

\subsubsection{Specific analytic and numerical calculations}

The assumptions for the analytic proofs also apply to the specific analytic and numerical calculations that we have done.
For those calculations, in addition it was necessary to choose specific equations of state for the dark energy.
The equation of state we chose is given in Eq.~\eqref{eq-state-de}. It has three parameters.  The classical Einstein equations can be solved to determine which
ranges of values of these parameters result in spacetimes with
either a little rip, or a final singularity of type I, II, III, or IV.  The details are given in Sec. III.

The background field calculations that we do for massive conformally coupled scalar fields involve the computation of
the full renormalized energy density in a specific spacetime containing a type III singularity.
This is done for fields with various masses, including $m = 0$.  For each mass the field is in a particular
fourth order adiabatic state.  This is described in detail in Appendix B.  We find that at late times the energy density approaches the value it has in the massless case.  This is the same type of calculation with the same result as that done for the massive conformally coupled scalar field in~\cite{ba}.  A different type of calculation was done in~\cite{bbfhd}
for this field in a spacetime with a type II singularity.  The result was that particle production effects are not important as the singularity is approached.
  These results for type I and III singularities (with supporting evidence from the results for type II singularities) provide evidence that massive conformally coupled scalar fields  become effectively conformally invariant near final singularities in the sense that the terms in the energy density which depend on the mass are subdominant as the singularity is approached.  This means we can treat them as massless, so the number of effectively massless fields near a singularity can be large.

As discussed above we investigate the effects of backreaction on various types of final singularities by solving the semiclassical backreaction equations when conformally invariant quantum fields are present.  The dimensionless parameter $\alpha$, which multiplies terms in the semiclassical backreaction equations due to both the quantum fields and the $R^2$ term in the gravitational Lagrangian, must be of order $10^9 $ in order for Starobinsky inflation to occur~\cite{huang}.
This large value, along with the large number of effectively massless quantum fields present, implies that below the Planck scale quantum effects due to the gravitational field can be ignored to leading order.

Having a large value for $|\alpha|$ and a large number of effectively conformally invariant fields near the final singularity means that quantum effects
will be important at much lower values of the spacetime curvature than was found in~\cite{ba}, where the effects of a single scalar field were considered.  In
their response~\cite{ba-response} to a comment on their paper~\cite{comment} the authors of~\cite{ba} pointed out that the backreaction solutions found in~\cite{comment},
in which the big rip singularity is avoided, vary on time scales comparable to or less than the Planck scale.  However, for values of $|\alpha|$ comparable to those
necessary for Starobinsky inflation, the types of solutions found in~\cite{comment} will vary on time scales substantially longer than the Planck scale.

Even though a large value of $|\alpha|$ results in solutions for which quantum effects are important on scales well below the Planck scale, there are still at least two potential
problems.  One is that the semiclassical approximation may become invalid at scales well below the Planck scale.  This has been discussed~\cite{flanagan-wald} in the context of a large $N$ expansion, with $N$ the number of identical quantum fields.  The main reason is that in the effective field theory approach~\cite{donoghue} there is
an infinite series of higher order terms in the gravitational Lagrangian, and this expansion is generally thought to break down when these terms become comparable to each other.
Of course it is possible that the coupling constants for the other terms are very small compared with $|\alpha|$.  So there might be a region where that term is large and the others are still small.

The second potential problem is that the presence of an $R^2$ term in the gravitational Lagrangian and the presence of quantum fields results in the appearance of higher derivative terms in the semiclassical backreaction equations.  These terms lead to a much larger number of solutions than occur for classical general relativity.  In many cases they also lead
to solutions which may follow a solution to the classical Einstein equation for some time but are unstable and eventually deviate substantially from it, usually by going into a period
of extremely rapid expansion or contraction.
It has been shown~\cite{simon,parker-simon} that in cosmology such solutions can be eliminated if one follows a procedure called
order reduction, in which the semiclassical backreaction equations are reduced to second order equations.  This is not always desirable.
For example, if order reduction is used, then Starobinsky inflation does not occur~\cite{simon}.

We study numerically the behaviors of the order-reduced equations in special cases and compare them to numerical solutions to the full semiclassical backreaction equations in those cases.
We find that, once quantum effects become important, a solution to the order-reduced equation generally deviates significantly from the corresponding solution to the
full semiclassical backreaction equations and that if the order-reduced solutions are continued into the regions
near the singularity, in many cases they have a qualitatively different effect on that singularity than the solutions to the semiclassical backreaction equations.
Thus solutions to the order-reduced equations are often not very useful for studying quantum effects near final singularities.

\subsubsection{Content summary}

In Sec.~\ref{sec:stress-tensor} we review the computation
of the stress-energy tensor for a quantized scalar field in a spatially flat Robertson-Walker spacetime.  Then we give the stress-energy
tensor for any conformally invariant free quantized field in these spacetimes and discuss the ambiguity in one of the renormalization
parameters.  We also discuss the semiclassical backreaction equations when conformally invariant fields are present and the way in which the higher derivatives can be eliminated using the method of order reduction.  In Sec.~\ref{sec:dark-energy-models} we discuss the models of the dark energy that we are using and the results when the classical Einstein equations are solved.  In Sec.~\ref{sec:background-field} we investigate quantum effects on the final singularities using the background field method, where the stress-energy tensor for the quantum fields and the terms coming from an $R^2$ term in the gravitational Lagrangian are evaluated in the classical spacetime geometry.  In Sec.~\ref{sec:possible-effects-sge} we discuss the types of semiclassical backreaction effects that can occur and the ways in which they can remove, avoid, or change the nature of a final singularity.  Sec. VI contains our backreaction results for the case $\alpha >0$, Sec. VII contains our backreaction results for $\alpha < 0$, and Sec. VIII contains our backreaction results for $\alpha = 0$.  Our results are summarized in Sec. IX.  In Appendix A the form we use for the stress-energy tensor for a massive conformally coupled scalar field is derived, and in Appendix B our method of choosing a state for this field is discussed.
Throughout we use units such that $\hbar = c = G = 1$, and our conventions are those of Misner, Thorne, and Wheeler~\cite{mtw}.

\section{Stress-energy tensor for quantum fields}
\label{sec:stress-tensor}

As mentioned in Sec.~\ref{introduction}, the effects of quantum fields on final singularities can be investigated using
background field calculations of the stress-energy tensor and by solving the semiclassical backreaction equations.
In this section we review the computation of the stress-energy tensor for conformally coupled massive scalar fields and conformally invariant fields in
a spatially flat Robertson-Walker spacetime.  We also write down the semiclassical backreaction equations for such a spacetime when conformally invariant
fields along with an $R^2$ term in the gravitational Lagrangian are present.

In a spatially flat Robertson-Walker spacetime, the line element can be written in the form
\be ds^2 = - dt^2 + a^2(t) d\, \vec{x}^2 \;. \label{metric} \ee
For this metric, a massive conformally coupled scalar field can be expanded in terms of a complete set of modes
such that~\cite{b-d-book}
\bes \be \phi =  \frac{1}{a(t)} \int d^3 k \, \left[ a_{\vec{k}} e^{i \vec{k} \cdot \vec{x}} \psi_k(t)   + a^\dagger_{\vec{k}} e^{-i \vec{k} \cdot \vec{x}} \psi^*_k(t)  \right] \;. \ee
The mode functions $\psi_k$ are solutions to the equation
\be \frac{d^2 \psi_k}{d t^2} + H \frac{d \psi_k}{dt} + \omega_k^2 \, \psi_k = 0 \;, \label{psi-eq} \ee
with
\bea  H & \equiv &  \frac{\dot{a}}{a}  \;, \label{H-def} \\
   \omega_k^2 &\equiv& \frac{k^2}{a^2} + m^2   \;,  \label{omega-def} \eea \ees
where dots denote time derivatives.  The solutions are normalized using the Wronskian condition
\be \psi_k \, \dot{\psi}_k^{*}  - \psi_k^{*} \, \dot{\psi}_k  = \frac{i}{a}  \;. \label{wronskian} \ee

In a Robertson-Walker spacetime there are two unique components of the stress-energy tensor, and they are connected by the conservation equation.
The full renormalized stress-energy tensor was written in~\cite{a-e} in terms of a part that usually must be computed numerically along with a part that
is known analytically.
In Appendix A we use the results of that paper and show that for a conformally coupled massive scalar field the energy density can be written in the form
\bes \bea \la \rho_q \ra &=& \la T^q_{tt} \ra =  \frac{1}{4 \pi^2 a^2} \int_0^\infty dk \, k^2 \left[ |\dot{\psi}_k|^2 + \omega_k^2 |\psi_k|^2 - \frac{\omega_k}{a} \right]
 + \rho_a  \;,   \\
 \rho_a &=& \frac{1}{2880 \pi^2} \left( - \frac{1}{6} \, {}^{(1)\!}H_{tt} + {}^{(3)\!}H_{tt} \right) + \frac{m^2}{288 \pi^2} G_{tt} \;. \label{rho-a} \eea \label{rho-m} \ees
Here
\bes \bea G_{tt} &=& \frac{3\dot{a}^2}{a^2} = 3 H^2 \;,  \\
   ^{(1)\!}H_{tt} &=& -36 \ddot{H} H + 18 \dot{H}^2 - 108 \dot{H}  H^2 \;, \label{Htt}  \\
   ^{(3)\!}H_{tt} &=& 3 H^4 \;.
\eea \ees
Note that the analytic part, $\rho_a$, given here is not the same as the analytic part in~\cite{a-e}.

If $m=0$, the normalized positive frequency solution to the mode equation is
\be \psi_k = \frac{1}{\sqrt{2 k}} \exp\left[-i k \int^t dt'/a(t') \right] \;.  \label{psi-m-0} \ee
Substituting this into~\eqref{rho-m} gives
\be \la \rho_q \ra = \rho_a = \frac{1}{2880 \pi^2} \left( - \frac{1}{6} \, ^{(1)\!}H_{tt} + {}^{(3)\!}H_{tt} \right) \;. \ee

The massless conformally coupled scalar field is conformally invariant. In a spatially flat Robertson-Walker spacetime,
the stress-energy tensor is known analytically if the fields are in the conformal vacuum state~\cite{b-d-book}.  The energy density is
\bes \be  \la \rho \ra =  -\frc16 \alpha_q \, {}^{(1)\!}H_{tt} +  \beta_q \, {}^{(3)\!}H_{tt}  \;, \label{density-conformal} \ee
with
\bea  \alpha_q &=&  \frac{1}{2880\pi^2} \left(N_0 + 6N_{1/2} + 12N_1 \right) \;, \label{alphaq} \\
\beta_q &=& \frac{1}{2880 \pi^2} \left(N_0 + 11N_{1/2} + 62N_1 \right) \;. \label{betaq} \eea \label{rho-quantum} \ees
Here $N_0$, $N_{1/2}$, and $N_1$ are the numbers of scalar, spin $\frc12$, and spin 1 fields respectively.

In a general spacetime, for $\la T_{ab} \ra$  it is necessary to have an $R^2$ term and a Weyl squared term, $C_{abcd} C^{abcd}$,
in the gravitational Lagrangian~\cite{amm}.\footnote{Because of the Gauss-Bonnet theorem~\cite{b-d-book}, it is also possible to replace the Weyl squared term
with a Ricci squared term, $R_{ab} R^{ab}$.}  The tensors that result from the variations of these terms are
\bes \bea  ^{(1)\!}H_{ab} &=& - \frac{1}{\sqrt{-g}} \frac{\delta}{\delta g^{ab}} \int d^4 x \, \sqrt{-g} \, R^2 = - 2 g_{ab} \Box R + 2 \nabla_a \nabla_b R - 2 R R_{ab}
+ \frc12 g_{ab} R^2 \;, \label{H1-def} \\
^{(C)\!}H_{ab} &=& - \frac{1}{\sqrt{-g}} \frac{\delta}{\delta g^{ab}} \int d^4 x \, \sqrt{-g} \, C_{abcd} C^{abcd} = -4 \nabla^c \nabla^d C_{abcd} + 2 R^{cd} C_{abcd} \;.
\label{HC-def} \eea \ees

The action which leads to the semiclassical backreaction equations when classical matter and free quantum fields are present is of the form
\be S = S^c_m + \Gamma^q + \frac{1}{16 \pi} \int d^4 x \sqrt{-g} R  + \frac{1}{2} \int d^4 x \sqrt{-g} \left[ \alpha_0 R^2 + \gamma_0 C_{abcd} C^{abcd} \right]  \;, \label{action} \ee
where $S^c_m$ is the action for the classical matter fields, $\Gamma^q$ is the one loop effective action for the quantum fields, and $\alpha_0$ and $\gamma_0$ are
dimensionless coupling constants.  The semiclassical backreaction equations are obtained from~\cite{b-d-book}:
\be  - \frac{2}{\sqrt{-g}} \frac{\delta}{\delta g^{ab}} S = 0  \; , \ee
with the result that
\be G_{ab} = 8 \pi \left[ T^{c}_{ab} + \la T^q_{ab} \ra + \alpha_0 \, ^{(1)\!}H_{ab} + \gamma_0 \, ^{(C)\!}H_{ab} \right] \;. \label{sce-1} \ee
Here $T^{c}_{ab}$ is the stress-energy tensor for the classical matter, and $T^q_{ab}$ is the stress-energy tensor operator for the quantum fields.

In Secs.~VI - VIII we solve the semiclassical backreaction equations in the case that conformally invariant quantum fields plus classical matter in the
form of dark energy are present.  The justification for omitting other types of quantum fields is given below.  We focus on
solving the time-time component of the equations.  For these fields in a spacetime with the metric~\eqref{metric}, the contribution of the term proportional
to $\alpha_q$ in~\eqref{density-conformal} is of the same form as the term in the time-time component of~\eqref{sce-1} which is proportional to $\alpha_0$.
Thus the effective energy density of the quantum field is
\be \rho_{qe} = \la \rho_q \ra + \alpha_0 \, {}^{(1)\!}H_{tt}  \;.  \label{rhoqe-def} \ee
Then the time-time component of the semiclassical backreaction equations is
 \bea  H^2 &=& \frac{8\pi}{3} (\rho_{de} + \rho_{qe})  \nonumber \\
 &=&   \frac{8\pi}{3} \left[ \rho_{de} + \alpha \left(-36 \ddot{H} H + 18 \dot{H}^2 - 108 \dot{H} H^2  \right) + 18 \beta H^4 \right].
  \label{fq} \eea
Here $\rho_{de}$ is the energy density of the dark energy, and
\be \alpha \equiv \alpha_0 - \frc16\alpha_q  \;, \qquad \beta \equiv \frc16\beta_q \; .\label{alpha-beta-def} \ee
Because $\alpha$ is the effective coefficient of an $R^2$ term in the gravitational Lagrangian, it is the value of $\alpha$ rather than $\alpha_0$ which affects
the behaviors of solutions to~\eqref{fq}.
Note that there is no contribution to~\eqref{fq} from $^{(C)\!}H_{tt}$, because all components of the Weyl tensor vanish in a conformally flat spacetime.

As is seen from~\eqref{alphaq}, the value of $\alpha_q$ depends on the number and types of conformally invariant fields present.  However,
there is no fundamental way to fix the value of $\alpha_0$ without invoking some theory of quantum gravity, and since  the two contributions combine together, this applies to the value of $\alpha$ as well. An experimental bound is $|\alpha| \lesssim 10^{74}$~\cite{stelle}. There is a much stronger bound if Starobinsky inflation occurred.  In this case for the density perturbations in the early universe to have the correct size it is necessary that $\alpha \sim 10^9 $~\cite{huang}.

As shown in~\eqref{betaq} and \eqref{alpha-beta-def}, the value of $\beta$ depends only on the number and types of conformally invariant fields present. Since conformally invariant
fields are massless, there is technically only one such field today, the electromagnetic field.  However, conformally coupled massive scalar fields along
with massive fields of spin $\frac12$ and $1$ are effectively conformally invariant if they are relativistic, as they are in the early universe if the temperature
is much larger than the mass, and if interactions can be neglected.  In the late universe, even near the singularity, the fields are not relativistic.  However,
if the contribution of the mass terms in the energy density is small compared with the massless terms, then the field is effectively conformally invariant.
This has been shown to be the case for the conformally coupled massive scalar field in some specific spacetimes with big rip singularities~\cite{ba}.  It is
also shown below for a specific spacetime with a type III singularity.  Thus it is very likely that if interactions don't contribute significantly to the energy density at late times,
then the massive fields that are present become, from the point of view of their energy densities, effectively massless, and thus effectively conformally invariant near the final singularity.  The standard model alone has $N_0 = 4$, $N_{1/2} = 45$, and $N_1 = 12$, and using these values in~\eqref{betaq} and \eqref{alpha-beta-def} shows that $\beta = \frac{1243}{17280 \pi^2}$.  Models such as supersymmetry and grand unified theories have many more fields.
Thus the value of $\beta$ for our universe is also unknown and could be relatively large, although probably not nearly as large as the value of $\alpha$ if Starobinsky
inflation occurred.

The effect of such large values of $\alpha$ and $\beta$ is to make it possible for significant backreaction effects due to $\rho_{qe}$ to occur on a scale well below the Planck scale, as they must for Starobinsky inflation to be viable.  Such effects could significantly alter the expansion of the universe near a final singularity.

One of the issues in solving the semiclassical backreaction equations involves the existence of higher derivative terms.  In particular, the time-time component~\eqref{fq} of the equations has up to three time derivatives of the scale factor, while the corresponding classical Einstein equation~\eqref{c-tt-eq} has only one.  This means that there are many more solutions to the equations, and that one must fix the starting value not only of the scale factor $a$ but also of its first two time derivatives.  There are different ways that have been proposed to deal with this problem.

One way is to eliminate the higher derivatives using a method called order reduction.  This has been developed for cosmological spacetimes by Parker and Simon~\cite{parker-simon}.  Here we use an approach that is equivalent to theirs.
One begins by using the classical equation~\eqref{c-tt-eq} and its derivatives to obtain expressions for $H$ and its derivatives in terms of the dark energy
density $\rho_{de}$ and its derivatives.  These are then used to compute the terms in the effective energy density for the quantum fields~\eqref{fq}, and the result is used
to obtain a new expression for $H$ on the left hand side of~\eqref{fq}.  The result is
\be
H^2 = \frac{8 \pi}{3} \left[ \rho_{de} +288 \pi^2 \alpha \left(\rho_{de} + p_{de}\right) \left(\rho_{de}+p_{de} - 4 \rho_{de} \frac{d p_{de}}{d \rho_{de}} \right) + 128 \pi^2 \beta \rho^2_{de} \right] \;.
\label{parker-simon-eq}  \ee
This is then integrated to obtain $a(t)$.

\section{Models for the dark energy}
\label{sec:dark-energy-models}

Our goal in this paper is to investigate the effects of quantum fields and an $\alpha_0 R^2$ term in the gravitational Lagrangian on
spacetimes with little rips and final singularities of types I - IV.  To do so we need to first consider solutions to the classical Einstein equations with the dark
energy as a source.  At late times in such universes the dark energy will dominate over all forms of classical matter as well as classical radiation and a cosmological constant if one is present.  Thus we include no other sources for the classical Einstein equations.

We model the dark energy as a perfect fluid with equation of state $p_{de} = p_{de}(\rho_{de})$.  In a spacetime with metric~\eqref{metric}
the conservation equation for the dark energy is
\be \label{conserve}
d \rho_{de} = -3 \left(\rho_{de}+p_{de}\right) \frac{da}{a} = -3H \left(\rho_{de}+p_{de}\right) dt \;.
\ee
Integrating this equation allows one to find the energy density $\rho_{de}$ as a function of the scale factor $a$ for any given equation of state.

\subsection{Models used for the analytic proofs}

 For the proofs in Secs. VI and VII regarding the behaviors of solutions to the semiclassical backreaction equations, we consider all equations of state which lead to a particular type of singularity when the classical Einstein equations are solved with the dark energy as a source.  In that sense these proofs are very general.

For little rips and for big rip (type I) singularities, $\rho_{de} \rightarrow \infty$ in the limit that $a \rightarrow \infty$.  The difference is that in the little rip
case this happens at an infinite proper time in the future, and in the big rip case it happens at a finite proper time in the future.  While one could go beyond this condition and distinguish between big and little rips, it is not necessary for the proofs in Secs. VI and VII.  There we simply consider all equations of state for which $\rho_{de} \rightarrow \infty$ in the limit that $a \rightarrow \infty$.

For the rest of the singularities we consider, the scale factor has a finite value at the singularity.
For type III singularities $\rho_{de} \rightarrow \infty$ as the singularity is approached.  For type II singularities, $\rho_{de}$ is finite at the singularity, but $p_{de}$ diverges as the singularity is approached.  Finally, for type IV singularities both $\rho_{de}$ and $p_{de}$ are finite at the singularity but $d p_{de}/da$ diverges as the singularity is approached.
These along with the fact that the dark energy is a perfect fluid are the only properties that we use for our proofs of the behaviors of solutions to the semiclassical backreaction equations in Secs. VI and VII.

\subsection{Models used for specific analytic and numerical calculations}

In Sec. IV we numerically compute the energy density for conformally coupled massive scalar fields in a background spacetime with a type III singularity.
In Secs. VI and VII we do specific analytic and numerical calculations to illustrate the analytic proofs of the behaviors of solutions to the semiclassical backreaction equations.
We also numerically solve the order-reduced semiclassical equations and compare the solutions to those of the full equations.

To do these calculations it is necessary to have specific equations of state for the dark energy.
We adopt a model for the dark energy with the equation of state
\be p_{de} = - \rho_{de} - A|\rho_{de} - \rho_s|^B  \;. \label{eq-state-de} \ee
Here $A$, $B$ and $\rho_s$ are parameters which can be varied to give different types of late time behaviors for the universe.  This equation of state is identical to the equations of state considered in~\cite{not},\footnote{Note, however, that their coefficients have different names: our coefficient $B$ corresponds to their $\beta$, and our $A$ is called either $B$ if $\rho_s = 0$ or $C$ if $\rho_s >0$.} save for the introduction of the absolute value, which allows us to continue the evolution of the universe beyond the singularity when $\rho_s >0$.  This model in the case $\rho_s=0$ was also considered as a driver for inflation in~\cite{barrow-inflation}.

Interestingly, there is a scaling symmetry for this particular form of the equation of state which connects solutions to~\eqref{fq} with large values of $\alpha$ and $\beta$ to those with smaller values.
If the various parameters in the equation along with the time are scaled so that
\be\label{rescale}
\left( \alpha, \beta, \rho, \rho_s, A, B , t \right) \rightarrow \left( N\alpha, N\beta, \rho/N, \rho_s/N, A N^{1-B}, B , t\sqrt{N} \right) \; ,
\ee
for some constant $N$, then for every solution to~\eqref{fq} with the original values of these variables there is a corresponding solution with the same type of behavior with the scaled values of these variables.  This scaling is very useful for numerical work.

Note that for $B=1$ and $\rho_s = 0$ we have an equation of state $p_{de} = w \rho_{de}$, with $w = -1-A$, which can represent dust ($A=-1$), radiation ($A=-\frc43$), or a cosmological constant ($A=0$). Since we are interested in future singularities, we restrict our attention to $A >0$.  We also restrict
our attention to $\rho_s \ge 0$.  As shown below, for $\rho_s = 0$ future singularities of type I, type III, or little rip cosmologies result for all values of $B$.  If $\rho_s >0$,
then future singularities of type II occur for $B<0$ and type IV for $0< B < \frac12$.
Finally, it should be emphasized that we are only concerned with the behavior of the universe near the future singularity.  For example, the case $\rho_s = 0$ and $B < 1$
is not expected to apply all the way to $\rho_{de}=0$ in a realistic model of the universe.

After finding $\rho_{de}$ as a function of the scale factor $a$ using the conservation equation~\eqref{conserve}, the result can be substituted into the time-time component of Einstein's equations,
\be H^2 = \frac{8 \pi}{3} \rho_{de} \; , \label{c-tt-eq} \ee
to find the behavior of the scale factor as a function of time.  The values of the parameters that lead to various types of final behaviors for the universe are discussed in detail in the following subsections and the results are summarized in Table I.

\subsubsection{Big rip singularities}
\label{sec:classical-big-rip}

We'll begin with the case $\rho_s = 0$.  For $B = 1$,
\be \rho_{de} = k a^{3 A}  \;, \label{rho-w-1} \ee
with $k$ a positive constant. Solving~\eqref{c-tt-eq}, one finds
\bes
\bea
a &=& \left[A\sqrt{6 \pi k} (t_s-t)\right]^{-\frac{2}{3A}} \; , \label{ac2} \\
\rho_{de} &=& \frac{1}{6\pi A^2 } (t_s-t)^{-2} \; .\label{rc2}
\eea
\ees
Since the scale factor and its derivatives diverge at a finite proper time and the energy density does as well, this is a big rip or type I singularity.

For $\rho_s = 0$ and $B \ne 1$,
\be\label{rhoscale1}
\rho_{de} =\left[3A(1-B)\log(a/a_s)\right]^\frac{1}{1-B} \;,
\ee
where $a_s$ is a positive constant.
 Substituting into~\eqref{c-tt-eq} one finds that
\bes
\bea
a &=& a_s \exp\left\{\left(3^BA\right)^\frac{1}{1-2B} \frac1{1-B} \left[(2B-1)\sqrt{2\pi}(t_s - t)\right]^\frac{2(B-1)}{2B-1}\right\} \; , \label{ac1} \\
\rho_{de} &=& \left[A(2B-1)\sqrt{6\pi} (t_s-t)\right]^{ \frac{-2}{2B-1}} \; .\label{rc1}
\eea \label{ac1-rc1}
\ees
Note that if $ \frc12 < B < 1$, the exponent $\frac{2B-2}{2B-1}$ is negative, resulting in a big rip singularity.

\subsubsection{Little rip}
\label{sec:classical-little-rip}

If $ B < \frc12 $, in~\eqref{rc1} then there is no future singularity at time $t = t_s$ because $2 B-1$ is negative.  Instead the divergence
occurs in the limit $t \rightarrow \infty$, so the universe ends with a little rip.
For the special case $B = \frc12$ it is straightforward to show that there is again a little rip.

\subsubsection{Type III singularities}
\label{sec:classical-specific-type-III}

For $B>1$, it can be seen from~\eqref{rc1} that in the limit $t \rightarrow t_s$, $a \rightarrow a_s$ and that $\rho_{de}$ diverges.  Thus the universe ends with a type III singularity.

\subsubsection{Type II and IV singularities}
\label{sec:classical-specific-type-II-IV}

If $\rho_s >0$, then
\be\label{rhoscale2}
\rho_{de} = \rho_s - {\rm sgn}(a_s-a) \left[ 3A(1-B) |\log(a_s/a)| \right]^\frac{1}{1-B} \; ,
\ee
where $a_s$ is a positive constant and ${\rm sgn}(a_s-a) = \pm1$, depending on the sign of $a_s-a$.
If $B<0$, there is a divergence in $d \rho_{de}/da$ and hence in $p_{de}$ at $a = a_s$, giving a type II singularity.
For $0<B<\frc12$, the divergence is in $d^2 \rho_{de}/da^2$, which yields a divergence in $d p_{de}/da$ resulting in a type IV singularity.
Less divergent singularities can be obtained for $\frc12 < B<1$. Substituting~\eqref{rhoscale2} into~\eqref{c-tt-eq} and solving near the singularity, one finds
\bes
\bea
a &\approx& a_s \exp \left\{ \sqrt{\frc83\pi\rho_s}(t-t_s) + \frac{1}{4-2B} \left[(3 \rho_s)^{B/2} A\right]^\frac{1}{1{-}B} \left[ (1{-}B)\sqrt{8\pi} | t-t_s|\right]^\frac{2{-}B}{1{-}B} \right\} \;, \\
\rho_{de} &\approx& \rho_s + \hbox{sgn}(t-t_s) \left[A(1{-}B)\sqrt{24\pi\rho_s}\left|t-t_s\right|\right]^\frac{1}{1-B} \; .\label{rc4}
\eea
\ees
Here the singularity is at $t = t_s$.  As mentioned above, the absolute value sign in~\eqref{eq-state-de} allows us to integrate~\eqref{c-tt-eq} through the singularity.
{\center
\begin{table}
\begin{tabular}{l c c}
\hline\hline
Classical classification & \multicolumn{2}{c}{~~~~~Equation of state parameters} \\ \hline
Little rip & ~~~~~~~~~~$\rho_s=0$ & $B \le \frac{1}{2}$ \\
Big rip/type I & ~~~~~~~~~~$\rho_s=0$ & $\frac{1}{2} < B \le 1$ \\
Type III & ~~~~~~~~~~$\rho_s=0$ & $1 < B $ \\
Type II  & ~~~~~~~~~~$\rho_s>0$ & $B < 0$ \\
Type IV & ~~~~~~~~~~$\rho_s>0$  &  $0 < B < \frac{1}{2}$\\ \hline\hline
\end{tabular}
  \caption{ Values of the parameters for our particular equation of state~\eqref{eq-state-de} that are used in the specific analytic and numerical calculations in Secs.~IV, VI, and VII.}
  \label{restab}
\end{table}
}

\section{Background field calculations}
\label{sec:background-field}

As discussed in the Introduction, one way to investigate the effects of quantum fields on final singularities is to evaluate the stress-energy tensor for the fields in the background spacetime containing the singularity.  We do so here first for the conformally coupled massive scalar field in a spacetime with a type III singularity and then for conformally invariant fields in spacetimes with all types of singularities.

\subsection{Massive conformally coupled scalar field}
\label{sec:background-field-massive}

Investigations of the effects of massive scalar fields on type I and type II singularities were done in~\cite{ba} and~\cite{bbfhd} respectively.  Here we show the results of the computation of the energy density for a massive conformally coupled scalar field in a spacetime with
a type III singularity.  The equation of state we chose for the dark energy is~\eqref{eq-state-de} with $A = 10$, $B = \frc54$, and $\rho_s = 0$.  With this
choice, Eqs.~\eqref{ac1-rc1} become
\bes \bea
a &=& a_s \exp \left\{ \frac{-2 (2 \pi)^{1/6} (t_s - t)^{1/3}}{3^{1/2} \, 5^{2/3}} \right\}  \\
\rho_{de} &=& \left[15 \sqrt{6 \pi} (t_s-t)\right]^{-4/3}  \;. \eea \ees
We also chose $a=1$ at $t=0$ and $a_s=10$, which implies $t_s \approx 79.1$.  The state for the field is a fourth order adiabatic vacuum state
which is obtained by setting the initial values equal to those for a fourth order WentzelKramersBrillouin, WKB, approximation at time $t=0$~\cite{b-d-book}.  The details are discussed in
Appendix B.

In Fig.~\ref{fig:massive} our results are shown for $m = 0$, $\frc14$, $\frc12$, $1$, and $2$.
It is clear that, for the range of times shown, the energy density for the massless field is always larger than that for a massive one.  However, it
is also clear that as the singularity is approached the energy density for a massive field approaches that for the massless one.  Thus the field
becomes effectively conformally invariant as the singularity is approached in the sense that its energy density approaches that of the massless scalar field.
This is exactly the same type of behavior as was found in~\cite{ba} for big rip singularities.  The investigation in~\cite{bbfhd} of the effects of particle production for
this field in spacetimes with type II singularities provides evidence that something similar will happen in this case as well.   As a result, we have evidence that for type I-III singularities the energy density for a conformally coupled massive scalar field always approaches that for the conformally invariant scalar field as the singularity is approached.
\begin{figure}[h]
\begin{center}
\includegraphics[angle=90,totalheight=0.3\textheight]{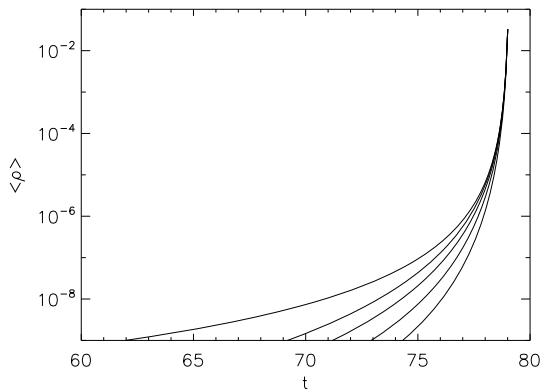}
\caption{Energy density for a conformally coupled scalar field.  From top to bottom the curves correspond to the
    cases $m = 0$, $\frac14$, $\frc12$, $1$, and $2$.}
\label{fig:massive}
\end{center}
\vspace{-5mm}
\end{figure}

\subsection{Conformally invariant fields}
\label{sec:background-field-massless}

We next want to evaluate the quantity $\rho_{qe}$ in~\eqref{rhoqe-def} for conformally invariant fields.
Since the values of $\alpha$ and $\beta$ are unknown, it is useful to separately consider the behaviors of the terms they multiply, which
we'll denote as $\rho_\alpha$ and $\rho_\beta$.  From~\eqref{fq} one finds
\bes \bea \rho_\alpha &=& -36 \ddot{H} H + 18 \dot{H}^2 - 108 \dot{H} H^2 \;, \label{rho-alpha-1} \\
     \rho_\beta &=& 18 H^4  \;. \label{rho-beta-1} \eea \ees
We can use~\eqref{c-tt-eq} and \eqref{conserve} to write these in terms of $\rho_{de}$ and its derivatives, with the result that
\bes \bea  \rho_\alpha &=& 288 \pi^2 \left(\rho_{de} + p_{de}\right)\left(\rho_{de} + p_{de} - 4 \rho_{de} \frac{dp_{de}}{d \rho_{de}} \right) \;, \label{rho-alpha-2} \\
      \rho_\beta &=& 128 \pi^2 \rho^2_{de}  \;.  \label{rho-beta-2} \eea \ees

Since $\beta >0$, it is clear for big rip, little rip, and type III singularities, where $\rho_{de}$ diverges at the singularity, that $\rho_\beta$ does so as well.
This term will become important in the semiclassical backreaction equations when
\be  \rho_{de} \sim \frac{1}{128 \pi^2 \beta} \;, \ee
which for values of $\beta$ of order unity is well below the Planck scale, where $\rho_{de} \sim 1$.  However, it is quite possible for this term to not be important near the
singularity for type II and type IV singularities.

The expression for $\rho_\alpha$ is more complicated because it depends on the derivatives of $\rho_{de}$.  Several terms in~\eqref{rho-alpha-2}
diverge for big rip, little rip, type III, and type II singularities.  For type IV singularities only the term proportional to $dp_{de}/d\rho_{de}$ diverges at the singularity.  In general it is clear that, unless there are significant cancellations between the diverging terms,
 $\rho_\alpha$ should become comparable to $\rho_{de}$ before the Planck scale is reached if $|\alpha| \gtrsim 1$, as is the case for Starobinsky inflation, where $\alpha \sim 10^9$.

\section{Possible effects of semiclassical gravity on final singularities}
\label{sec:possible-effects-sge}

 Before discussing the details of the different late time behaviors, it is useful to first discuss whether and how backreaction effects in semiclassical gravity can remove, avoid, or change the nature of a final singularity.

 The key point relating to the possible removal or avoidance of final singularities is the behavior of the dark energy.  As discussed in Sec.~III.A, the fact that the dark energy is a perfect fluid means that its density, $\rho_{de}$, and hence its pressure are a function of the scale factor, and the form of that function is determined by the equation of state.  Hence if there is a divergence in, for example, $p_{de}$, but not $\rho_{de}$ at some particular value of the scale factor, $a = a_s$, then this will occur for any spacetime in which the scale factor reaches the value $a_s$ regardless of whether the scale factor is a solution to the classical Einstein equations or the semiclassical Einstein equations or a solution to neither.  Therefore it is not possible for final singularities to be removed in semiclassical gravity if the dark energy is a perfect fluid.

 However, it is possible to avoid a little rip or a type I-IV singularity.  This will happen if the scale factor never reaches the value at which the singularity occurs.  There are only two ways in an expanding universe that the scale factor will never reach some particular value $a_s$.  Either there is a bounce at which the scale factor reaches a maximum size which is less than $a_s$ or there is a different singularity at a value of the scale factor which is less than $a_s$.  In both cases the original singularity is avoided, although in the second case it is replaced by another singularity.  In the first case whether or not the original singularity is replaced by another singularity depends upon the behavior of the universe after the bounce.  In the cases that follow in which a bounce occurs we do not pursue the question of the subsequent evolution of the universe, as this involves making assumptions about the matter and radiation content and the states of the quantum fields and is thus beyond the scope of this paper.

Another possibility which can occur if the singularity is not avoided is that it can be either strengthened or softened.  This is a subtle point.  As discussed above, whether or not there is a divergence at a given value of the scale factor in $\rho_{de}$ or some quantity related to it, depends on the equation of state.  In general, the type of divergence determines, through the classical Einstein equations, which time derivatives of the scale factor diverge when the singularity is reached.  However, if the semiclassical backreaction equations are solved then there are other possibilities.  One is that a divergence in $\rho_{qe}$ in~\eqref{fq} or one of its derivatives at the singular value of the scale factor $a_s$ will be stronger than that of the dark energy and result in the strengthening of a singularity.
For example we shall show below that in some cases little rips are turned into big rip singularities, because $\rho_{qe}$ diverges much more rapidly than $\rho_{de}$ for the corresponding solutions to the semiclassical backreaction equations.

It can also work the other way.  A divergence in $\rho_{qe}$ at $a = a_s$ can result in a cancellation of the divergence in, for example, $\rho_{de}$ but no cancellation of the divergence in $d \rho_{de}/da$.  In this case a type III singularity is transformed to a type II singularity and the singularity is softened.  In some cases the singularity is softened to the point that the lowest order in which the divergence is not canceled is  either $d^3 \rho_{de}/da^3$ or $d^4 \rho_{de}/da^4$. In these cases we say that the singularity is effectively removed.

 Because it is possible for semiclassical effects to change the nature of a classical singularity, when solving the semiclassical backreaction equations, we classify the future singularity based on the behavior of the scale factor and its derivatives, and not on the behavior of $\rho_{de}$ and its derivatives.  Therefore if the scale factor and its derivatives diverge only in the infinite proper time limit we say that a little rip occurs.  If they diverge in finite proper time then a big rip or type I singularity occurs.
 If a solution to the semiclassical backreaction equations has, at the singularity, $a \rightarrow a_s < \infty$ and $H \rightarrow \infty$, then
it is a type III singularity.  If at $a_s$, $H$ is finite but $\dot{H}$ diverges then it is a type II singularity.  If instead both $H$ and $\dot{H}$ are finite at $a_s$ but $\ddot{H}$ diverges then it is a type IV singularity.  And if only some higher derivative of $H$ diverges then we say that the singularity is so soft that it has been effectively removed.

\section{Backreaction Effects for $\alpha > 0$}\label{sec:PartA}

In this section we consider the case $\alpha >0$, which is the sign of $\alpha$ necessary for Starobinsky inflation to occur.  As discussed in the Introduction, we
have analytic proofs which describe the behaviors of solutions to the semiclassical backreaction equations in cases where classical general relativity predicts the existence of little rips or type I - IV singularities.  We also have analytic and numerical calculations for specific values or ranges of values of the parameters in the equation of state~\eqref{eq-state-de}.  These are used to illustrate the behaviors in the proofs and also to investigate the solutions to the order-reduced equation~\eqref{parker-simon-eq}.
The pattern we shall follow for particular types of classical final behaviors is to first give the proof which describes the behaviors of solutions to the
semiclassical backreaction equations in that case and then to discuss the analytic and numerical calculations for specific examples.

First we give a simple proof which shows final singularities are never avoided if $\alpha > 0$.
The key element in the proof was pointed out in~\cite{fhh}.  If $\alpha > 0$, then all the terms on the right side of~(\ref{fq}) either vanish or are positive if $H = 0$, and therefore this equation cannot have $H=0$ at any time.   This means that the scale factor will monotonically increase until either a singularity is reached or it becomes infinite.  As discussed in Sec.~\ref{sec:possible-effects-sge}, we are modeling the dark energy as a perfect fluid, and thus $\rho_{de} = \rho_{de}(a)$.  Therefore in cases where classical little rips, big rip singularities, or type III singularities occur, it is impossible to prevent $\rho_{de}$ from diverging.  In cases where type II or type IV singularities occur it is impossible to prevent $p_{de}$ and/or $dp_{de}/da$ from diverging.  However, as shown below, there are cases in which the singularity is strengthened as well as cases where it is softened or effectively removed.

\subsection{Classical little rips and big rip singularities}

\subsubsection{Proof that only big rip singularities occur}

In this section we show that for all cases in which classically there is either a little rip or a big rip singularity,
 all solutions to the semiclassical backreaction equations~\eqref{fq} end in big rip singularities.

First it is useful to make the change of variables~\cite{fhh,rr}
\be   y = a^3 \; ,  \qquad  f = (a \dot{a})^{3/2}   \;. \label{y-f-def}  \ee
Then~\eqref{fq} takes the form
\be  \left(\frac{f}{y} \right)^{4/3} = \frac{8 \pi}{3} \left[\rho_{de} - 216 \alpha \frac{f^{5/3}}{y^{2/3}} \frac{d^2 f}{d y^2} + 18 \beta \left(\frac{f}{y}\right)^{8/3} \right] \;, \label{fq-fy} \ee
or
\be \frac{d^2 f}{d y^2} = \frac{y^{2/3}}{216 \alpha f^{5/3}} \rho_{de} + \frac{\beta}{12 \alpha} \frac{f}{y^2} - \frac{1}{576 \pi \alpha y^{2/3} f^{1/3}}  \;. \label{fq-2} \ee

For both little rips and big rips, $H \rightarrow \infty$ as $a \rightarrow \infty$.  Thus examination of~\eqref{fq} shows that the term on the left which is $H^2$ is much smaller
in this limit than the $\beta H^4$ term on the right.  To find the asymptotic behaviors of solutions one can neglect the former term, which is also the term on the left
in~\eqref{fq-fy} and the last term on the right in~\eqref{fq-2}.  In particular the latter equation then becomes
\be \frac{d^2 f}{d y^2} = \frac{y^{2/3}}{216 \alpha f^{5/3}} \rho_{de} + \frac{\beta}{12 \alpha} \frac{f}{y^2} \;. \label{fq-3} \ee
To study the solutions of~\eqref{fq-3}, let us define two new functions $c_\pm(y)$ by the simultaneous equations
\bes\bea f &=& c_+(y) y^{p_+} + c_-(y) y^{p_-} \; , \label{c-pm-def1} \\
\frac{df}{dy} &=& c_+(y) p_+ \; y^{-1+p_+} + c_-(y) p_- \; y^{-1+p_-} \; , \label{c-pm-def2} \eea
where $p_\pm$ are the roots of the equation $p^2-p=\beta/12\alpha$, namely
\be p_\pm = \frac12 \left(1 \pm  \sqrt{1+\frac{\beta}{3\alpha}} \right) \; . \label{p-def} \ee \ees
We note that $p_+ > 1$ and $p_- < 0$.

From \eqref{y-f-def} it is obvious that $f$ is positive, and if $a$ and $H$ are both becoming large, that $f$ is growing, so that $df/dy > 0$.  From~\eqref{c-pm-def1} and \eqref{c-pm-def2}, we can see that the only way both $f$ and its derivative can be positive is if $c_+(y) > 0$.  If we take the first derivative of \eqref{c-pm-def1} and compare it to \eqref{c-pm-def2}, we discover that
\be y^{p_+} \frac{d}{dy}c_+(y) + y^{p_-}\frac{d}{dy}c_-(y) = 0 \;. \label{c-pm-eq1} \ee
If we substitute \eqref{c-pm-def2} into \eqref{fq-3} and use the fact that $p_\pm^2-p_\pm = \beta/12\alpha$, we find
\be p_+y^{p_+} \frac{d}{dy} c_+(p) + p_- y^{p_-} \frac{d}{dy} c_- (p)= \frac{y^{5/3} \rho_{de}}{216 \alpha f^{5/3}} \; . \label{c-pm-eq2} \ee
Eliminating $c_-(y)$ from \eqref{c-pm-eq2} using \eqref{c-pm-eq1}, we find
\be \frac{d}{dy}c_+(y) = \frac{y^{5/3} \rho_{de}}{216\alpha f^{5/2} y^{p_+} \left(p_+ - p_-\right)} \; . \label{fq-6} \ee

The point of~\eqref{fq-6} is that $c_+(y)$ is an increasing function of $y$, and we already know it is positive.  As $y \rightarrow \infty$, it is clear from \eqref{c-pm-def1} that the $c_+$ term dominates, so we can approximate
 \be f \approx c_+(y) y^{p_{+}} \;.  \label{fapprox} \ee
   Substituting our definitions \eqref{y-f-def} as well as \eqref{p-def} into~\eqref{fapprox}, we can solve for $t$ to find
\be t = \int \frac{da}{a^{\sqrt{1+\beta/3\alpha}} \left[c_+\left(a^3\right)\right]^{2/3}} \; . \label{t-eq1} \ee
This integral converges as $a \rightarrow \infty$ for any rising function $c_+$, which means that the universe attains infinite size in a finite time.
Hence we conclude that for all cases in which classically
there is either a little rip or a big rip singularity, all solutions to the semiclassical backreaction
equations~\eqref{fq} end in big rip singularities.

If $|\rho_{qe}| \gg |\rho_{de}|$ as $a \rightarrow \infty$ then the first term on the right in~\eqref{fq-3} can be neglected and~\eqref{c-pm-def1}
becomes an exact solution to the resulting equation with $c_{\pm}$ constants.  In this case the asymptotic behavior of the scale factor can be obtained
by integrating~\eqref{t-eq1} with the result that
\be a \approx c (t_s-t)^{1/(1-\sqrt{1+\beta/3 \alpha})}  \;, \label{a-big-rip-1} \ee
where $c$ and $t_s$ are constants.  Since the exponent is negative, there is a big rip singularity at time $t = t_s$.

\subsubsection{Specific analytic and numerical calculations}

The results of the previous section are in agreement with results previously found for the equation of state $p_{de} = w \rho_{de}$ with $w$ a constant and $w < -1$.  Classically there is a big rip, or type I, singularity in this case and several authors~\cite{haro-2,comment,haro,hae-1} have pointed out that a singularity must remain.  However, they did not study the nature of that singularity.  We do so next and then generalize to other cases in which our specific equation~\eqref{eq-state-de} leads to classical little rips or big rip singularities.

As shown in Sec. III, if we set $\rho_s = 0$ and $B = 1$ then our specific equation of state is equivalent to $p_{de} = w \rho_{de}$ with $w = -1-A$.
To study the behavior of the solutions to the semiclassical backreaction equations~\eqref{fq-2} in this case we begin by assuming that $|\rho_{qe}| \gg |\rho_{de}|$  as $a \rightarrow \infty$ so that~\eqref{a-big-rip-1} gives the asymptotic behavior of the scale factor.  We can then evaluate $\rho_{qe}$ and $\rho_{de}$ to find out if this condition is
satisfied.  Substituting~\eqref{a-big-rip-1} into~\eqref{fq} one finds that the terms in $\rho_{qe}$ are all proportional to $(t_s-t)^{-4}$. Substitution into~\eqref{rho-w-1} gives
\be \rho_{de} \approx kc^{3A} (t_s-t)^{3A/(1-\sqrt{1+\beta/3\alpha})} \; . \ee
Comparing the powers, we find that $|\rho_{de}|$ is negligible compared to the individual terms in $\rho_{qe}$  if $3A < 4\left(\sqrt{1+\beta/3\alpha}-1\right)$, which is equivalent to $27A^2 \alpha + 72 A \alpha < 16 \beta$.
In this case~\eqref{a-big-rip-1} is a self-consistent solution to~\eqref{fq-2} near the singularity.  Since the classical behavior is a power law given by~\eqref{ac2} and the quantum behavior is a different power law given by \eqref{a-big-rip-1}, one sees that the singularity is softened if $27A^2\alpha + 36 A \alpha < 4\beta$ and strengthened if the inequality
is in the other direction.

If $27A^2 \alpha + 72 A \alpha \ge 16 \beta$, then it must be that our original assumption is wrong and therefore that as the singularity is approached $\rho_{de}$ is comparable to the terms in $\rho_{qe}$.  We find that if  $27A^2 \alpha + 72 A \alpha > 16 \beta$, the solution takes the form
\bes\bea a &\approx& \left[ \frac{32\left(27A^2\alpha+72A\alpha-16\beta\right)}{9A^4k(t_s-t)^4} \right]^{\frac1{3A}} \; , \label{ac5}\\
\rho_{de} &\approx& \frac{32\left(27A^2\alpha+72A\alpha-16\beta\right)}{9A^4(t_s-t)^4} \; .
\eea \ees
If $27A^2 \alpha + 72 A \alpha = 16 \beta$, then we find
\bes
\bea
a &\approx& \left[ \frac{16\alpha(3A+4)}{3A^3\,k\,(t_s - t)^4|\ln\left(t_s-t\right)|}\right]^\frac{1}{3A} \; ,\label{aq3} \\
\rho_{de} &\approx& \frac{16\alpha(3A+4)}{3A^3(t_s - t)^4|\ln\left(t_s-t\right)|} \; , \label{rq3}
\eea
\ees
In both cases, comparing with~\eqref{ac2} one can see that the singularity is strengthened.

It was shown in Sec. III that for the specific equation of state~\eqref{eq-state-de}, big rip solutions to the classical Einstein
equations occur if $\rho_s = 0$ and $\frc{1}{2} < B \le 1$ while little rip solutions occur if $\rho_s = 0$ and $B \le \frc{1}{2}$.
In general for $B < 1$ the behavior of $\rho_{de}$ is given by~\eqref{rhoscale1}.  Recall that if the condition $|\rho_{qe}| \gg |\rho_{de}|$
is satisfied then the scale factor near the singularity has the behavior~\eqref{a-big-rip-1} and $\rho_{qe}\sim (t_s-t)^{-4}$.  In this case
substituting~\eqref{a-big-rip-1} into~\eqref{rhoscale1} shows that the condition $|\rho_{qe}| \gg |\rho_{de}|$ is satisfied near the singularity
and thus that~\eqref{a-big-rip-1} gives the correct behavior of the scale factor near the singularity.  Comparison with the behavior~\eqref{ac1} of the scale factor
for the solution to the classical Einstein equation when $B < 1$ shows that the singularity is always weakened by quantum effects if $\frac{1}{2} < B < 1$.
For $B \le \frc{1}{2}$ there is classically a little rip which as we have just shown is turned into a big rip singularity.

\begin{figure}
\centering
\subfloat{\includegraphics[angle=90,width=3.1in,clip]{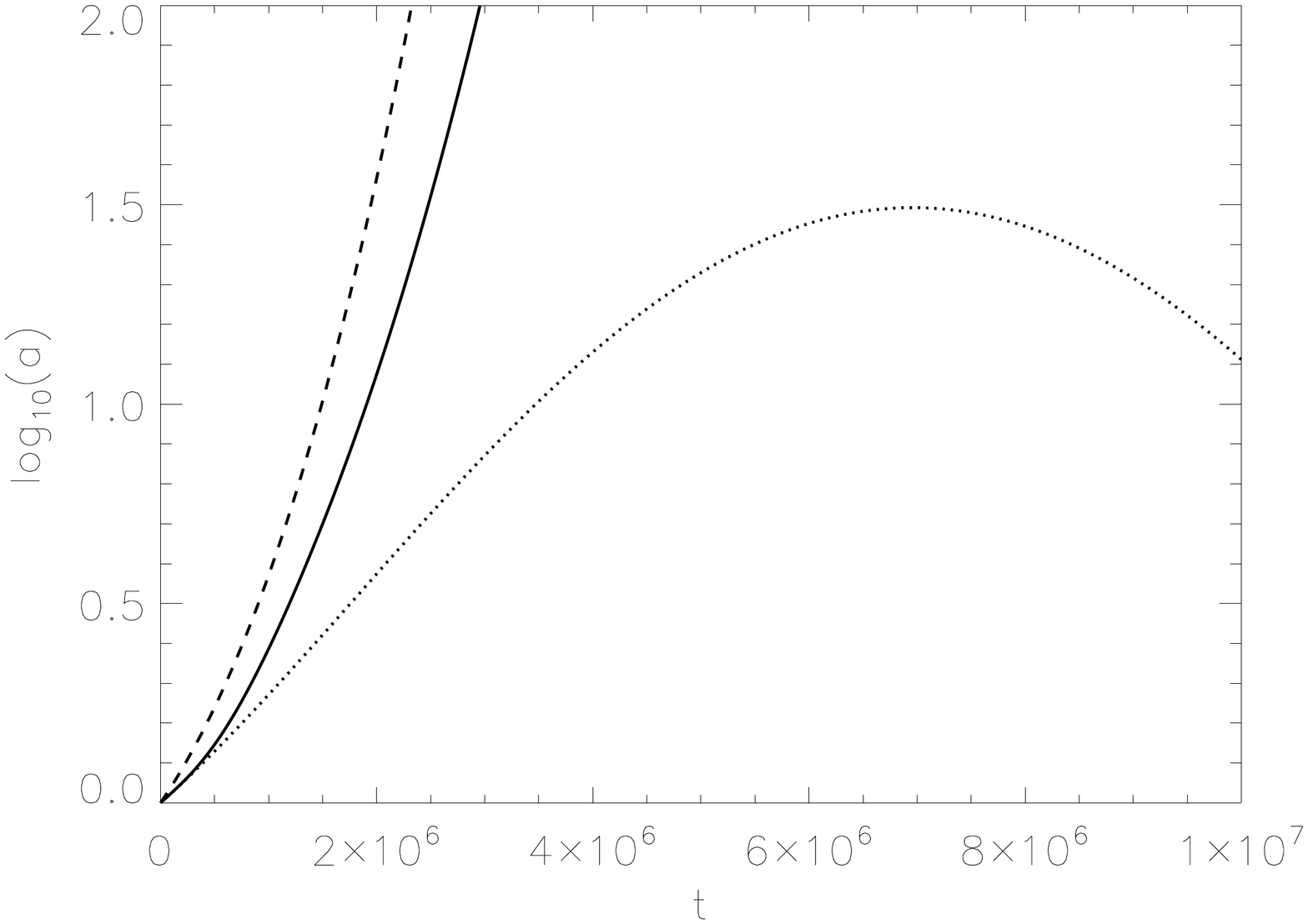}}\hfill
\subfloat{\includegraphics[angle=90,width=3.1in,clip]{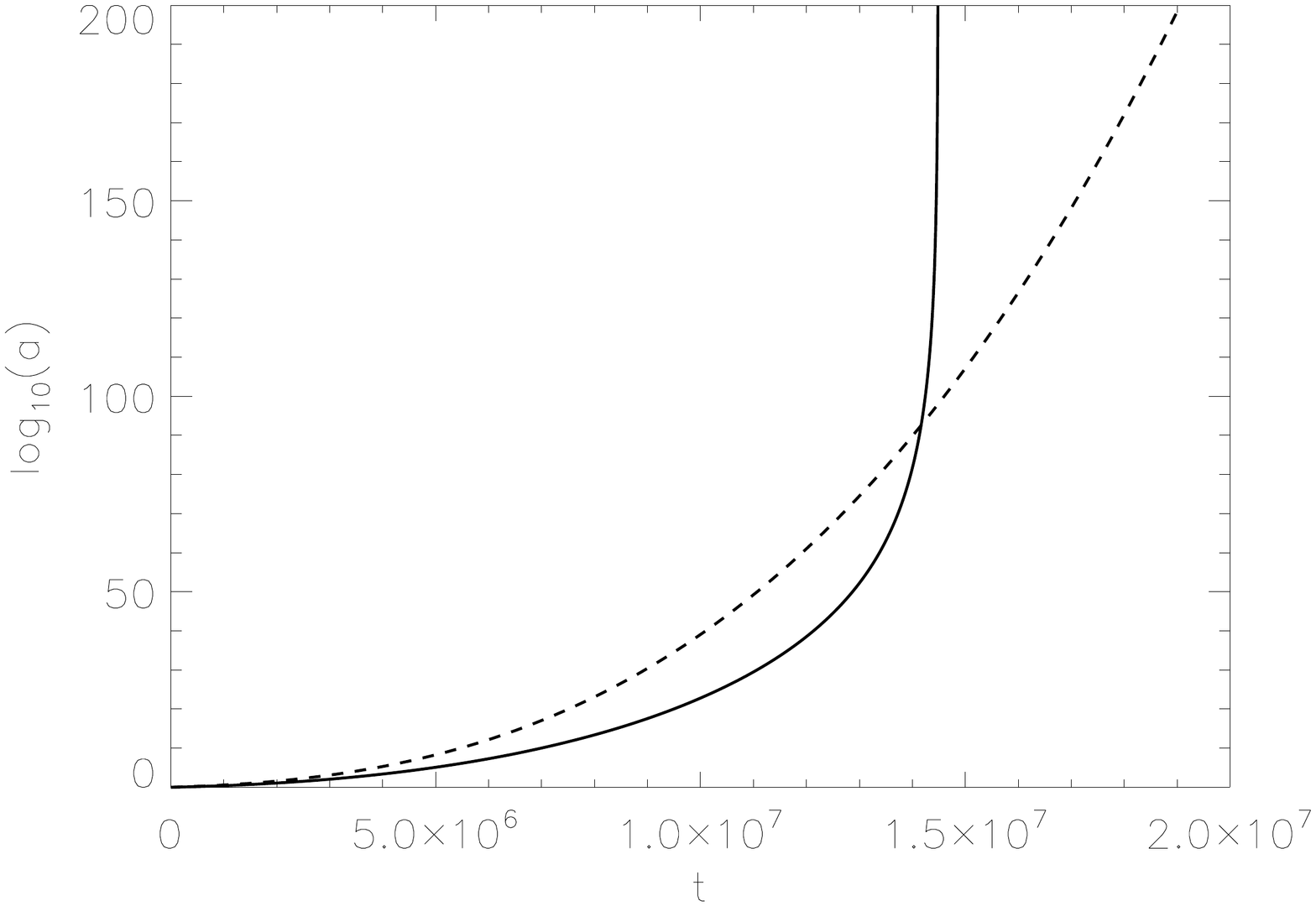}}
\caption{Scale factor $a(t)$ for $\alpha=10^9$, $\beta=10^8$ for a classical little rip model for the equation of state~\eqref{eq-state-de} with $\rho_s = 0$, $B=\frc14$, and $A=10^{-10}$. The dashed line denotes the classical
solution and the solid line denotes the solution to the semiclassical backreaction equation~\eqref{fq}.  In the plot on the left the dotted line denotes
the solution to the order reduced equation~\eqref{parker-simon-eq}.  Note that the solution to the semiclassical equation diverges rapidly, implying that it has a big rip singularity.
The solution to the order reduced equation undergoes a bounce, which is not possible for solutions to the full semiclassical equation~\eqref{fq} when $\alpha > 0$. }
\label{Little_Rip_Fig}
\end{figure}

We have numerically solved the semiclassical backreaction equation~\eqref{fq} using the model with equation of state~\eqref{eq-state-de} for the case $B=\frc14$, $A=10^{-10}$, which classically
gives a little rip.  Our results are shown in Figs.~\ref{Little_Rip_Fig}, where it can be seen explicitly that the classical little rip is converted into a big rip singularity.  For one of the plots we also include the solution to the order reduced equation~\eqref{parker-simon-eq}. As might be expected, when quantum effects are small, the solution to the order reduced equation is close to that of the exact equation~\eqref{fq}.  However, the behavior of the solution to the order reduced equation becomes very different when quantum effects become significant.  In fact in this case it has a bounce, which we have shown above is not possible for solutions to the exact semiclassical equation~\eqref{fq} if $\alpha >0$.

\subsection{Classical singularities of type III}
\subsubsection{Analytic proofs regarding solutions to the semiclassical backreaction equations}
\label{sec:alpha-gt-0-type-III-proofs}

If there is classically a type III singularity, then $\rho_{de}$ diverges in the limit $a \rightarrow a_s$ for some $a_s < \infty$.  Thus it is obvious that such a singularity cannot turn into either a big rip or a little rip.  Further, since $\rho_{de}$ diverges, at least one other term in~\eqref{fq} must diverge for any solution to this equation.  This means that the softest possible singularity that could occur is a type IV, for which the third derivative of the scale factor diverges but the lower order ones are finite.  We find that it is possible to have a type III singularity either stay a type III singularity or soften into a type II or type IV singularity.  What happens depends on how $\rho_{de}$ behaves as a function of the scale factor.

To see the conditions under which each behavior occurs, first note
   that regardless of whether a solution to the semiclassical backreaction equations~\eqref{fq-2} has a type II, III, or IV singularity, the third time derivative of the scale factor will diverge, and
hence so will $d^2 f/dy^2$.  Since $y$ approaches a constant near the singularity and $f$ either approaches a constant or diverges, the third term on the right hand side of~\eqref{fq-2}
cannot diverge at the singularity.  Suppose that the dominant term as the singularity is approached is the second one.  The solution is given by \eqref{c-pm-def1} with constant functions $c_\pm$, and has no divergence at finite $y$.  So this term cannot be the dominant one as the singularity is approached.  That means that near the singularity one must have
\be \frac{d^2 f}{d y^2} = \frac{y_s^{2/3}}{216\alpha} \frac{\rho_{de}}{f^{5/3}}   \;, \label{fq-5} \ee
where $y_s=a_s^3$.  Since there are no bounce solutions when $\alpha >0$ as the singularity is approached, $f$ either approaches a positive constant or diverges as $y \rightarrow y_s$.

First consider the case in which the solution to~\eqref{fq-2} yields either a type II or a type IV singularity.  In these cases  $f_s \equiv f(y_s)$ is finite.  Using this condition and integrating~\eqref{fq-5} once we find that near the singularity
\bea  \frac{df}{dy} &=& \frac{y_s^{2/3}}{216 \alpha f_s^{5/3}} I_1 \;,   \label{f-deriv-soln} \\
      I_1 &=&  \int^y dy_1 \, \rho_{de}(y_1)  \;.  \label{I1}    \eea
For a type IV singularity, $df/dy$ is finite at the singularity.  Thus the condition for solutions to the semiclassical backreaction equations to have a type IV singularity
is
\be   \lim_{y\rightarrow y_s} I_1 < \infty  \;. \ee

For a type II singularity, $df/dy$ diverges at the singularity.  So one condition for solutions to the semiclassical backreaction equations to have a type II singularity is
 \be   \lim_{y\rightarrow y_s} I_1 = \infty  \;. \ee
Integrating~\eqref{f-deriv-soln} once gives near the singularity
\bea  f &=& \frac{y_s^{2/3}}{216 \alpha f_s^{5/3}} I_2 \;,   \label{f-soln} \\
      I_2 &=&  \int^y dy_1 \,\int^{y_1} dy_2 \rho_{de}(y_2)  \;.  \label{I2}    \eea
For $f(y_s)$ to be finite it is necessary that
\be \lim_{y\rightarrow y_s} I_2 < \infty  \;. \ee

For a type III singularity, $f$ diverges at the singularity.  In this case, integrating~\eqref{fq-5} twice near the singularity yields
\be f = \frac{y_s^{2/3}}{216\alpha}  \int^y dy_1  \int^{y_1} dy_2 \frac{\rho_{de}(y_2)}{f(y_2)^{5/3}}  \;.  \label{f-soln-4}  \ee
For $f$ to diverge at the singularity it is necessary that
\be  \lim_{y \rightarrow y_s} I_2 = \infty \;. \ee
Note that this is also a sufficient condition because the analysis for type II singularities above shows
that $f(y_s)$ is finite if $I_2$ is finite at the singularity.

\subsubsection{Specific analytic and numerical calculations}
\label{num-alpha-gt-0-type-III}

We can apply these formulas to our specific models for classical type III singularities.  As shown in Sec.~\ref{sec:classical-specific-type-III}, classical type III singularities occur for the equation of state~\eqref{eq-state-de} if $\rho_s=0$ and $B>1$.
  Using~\eqref{rhoscale1}, we see that near the singularity, $\rho_{de} \sim (a_s - a)^{-1/(B-1)} \sim (y_s-y)^{-1/(B-1)}$.  Integrating twice, it is easy to see that \eqref{I2} diverges if $B \le \frc32$ while \eqref{I1} diverges if $B \le 2$.  Thus type III singularities remain type III for $1 < B \le \frc32$, convert to type II for $\frc32 < B \le 2$, and get softened further to type IV for $B > 2$.

For type III singularities with $1 < B < \frc32$, we find near the singularity, solutions to~\eqref{fq} can be expanded in powers of $t_s - t$, with $t_s$ the time at which the singularity is reached.  The result to leading order is
\be a \approx a_s \exp \left[ - c_\gamma (t_s - t)^\gamma \right] \;,  \label{a-sol-1} \ee
with
\bes \bea
\gamma &=&\frac{4B-4}{2B-1} \; , \label{gamma2} \\
c_\gamma &=& \gamma^{-1} \left[18\alpha (1-\gamma)(3-\gamma)\right]^\frac{1-B}{2B-1} \left[\frc34 A (2B-1)\right]^\frac{-1}{2B-1} \; , \label{cg1} \\
\rho_{de} &\approx& \left[3A(B-1)c_\gamma\right]^\frac{-1}{B-1} (t_s-t)^\frac{-4}{2B-1} \; .\label{rq4}
\eea
\ees

\begin{figure}
\centering
\subfloat{\includegraphics[angle=90,width=3.1in,clip]{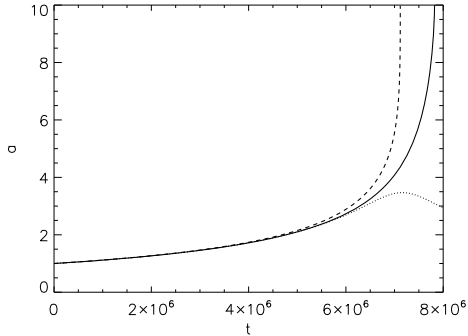}}\hfill
\caption{Scale factor $a(t)$ for a classical type III singularity with $\alpha=10^9$, $\beta=10^8$, $B=\frc54$ and $A=3000$.  The classical singularity occurs at $a = a_s = 10$.  The dashed line denotes the classical solution, the solid line denotes the solution to the semiclassical backreaction equation~\eqref{fq}, and the dotted line denotes
the solution to the order reduced equation~\eqref{parker-simon-eq}.  Note that the solution to the semiclassical backreaction equation remains a type III singularity since $\dot{a}$ diverges at the singularity.  However, the solution to the order reduced equations undergoes a bounce, which is not possible for solutions to the full semiclassical equation~\eqref{fq} when $\alpha > 0$.  }
\label{Type_III_1_Fig}
\end{figure}

Comparing~\eqref{ac1}, \eqref{a-sol-1} and \eqref{gamma2}, it is clear that $0< \frac{2B-2}{2B-1} < \gamma < 1$ for $1<B<\frc32$, so that $\dot a$ still diverges when including quantum effects, though as a weaker negative power of $t_s-t$.  But comparing~(\ref{rc1}) and (\ref{rq4}), the dark energy density rises as a more negative power of $t_s-t$. The reason is that the dark energy density is being partially canceled by the quantum energy density.  We will nonetheless consider this a weakening of the singularity because we will focus on the behavior of the scale factor $a$ and not the dark energy density.  The behavior of the scale factor for this type of model is exhibited in Fig.~\ref{Type_III_1_Fig}.  Note that, once again, the order reduction approach leads to a bounce, which cannot occur for solutions to the full semiclassical backreaction equation~\eqref{fq}.

\begin{figure}
\centering
\subfloat{\includegraphics[angle=90,width=3.1in,clip]{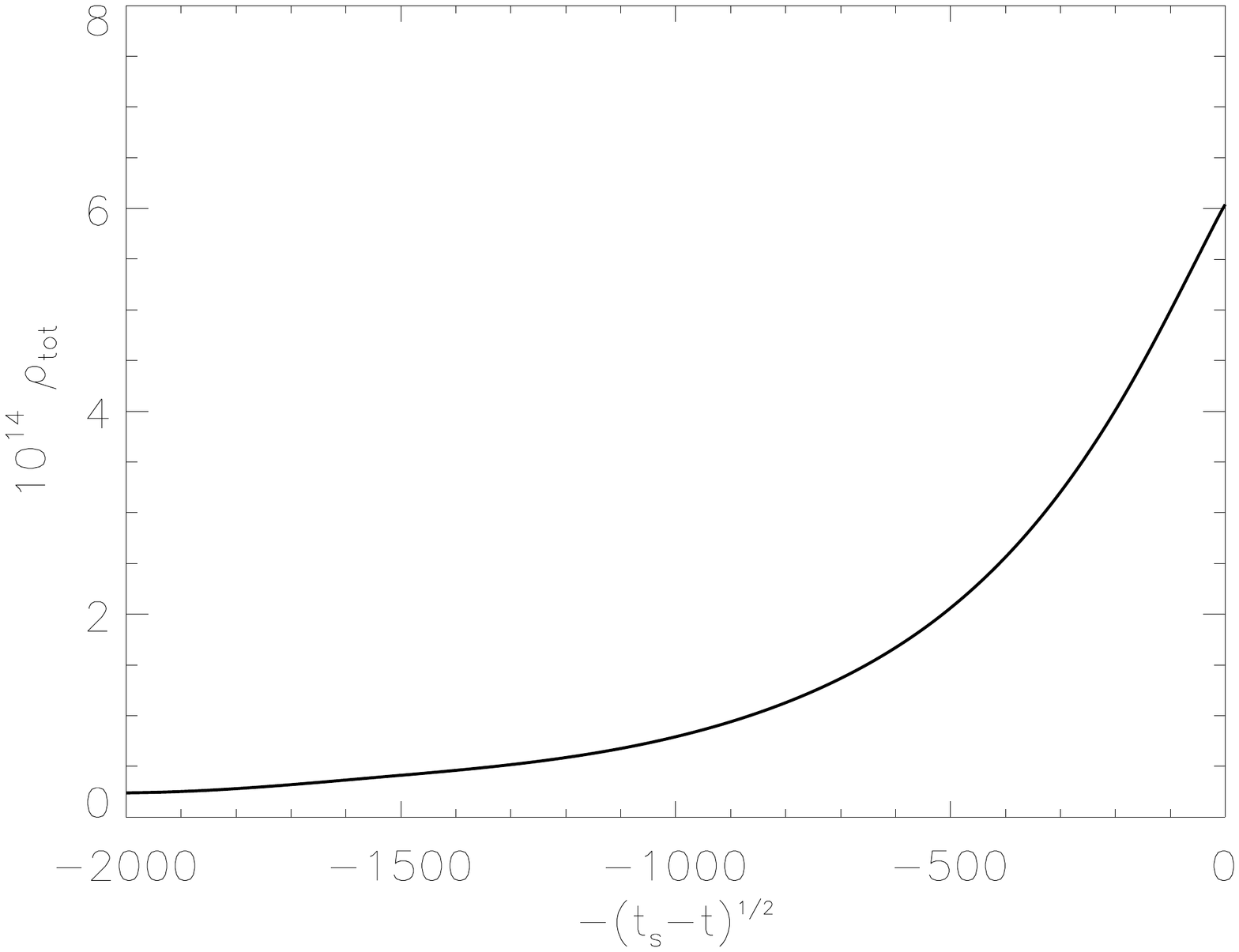}}\hfill
\subfloat{\includegraphics[angle=90,width=3.1in,clip]{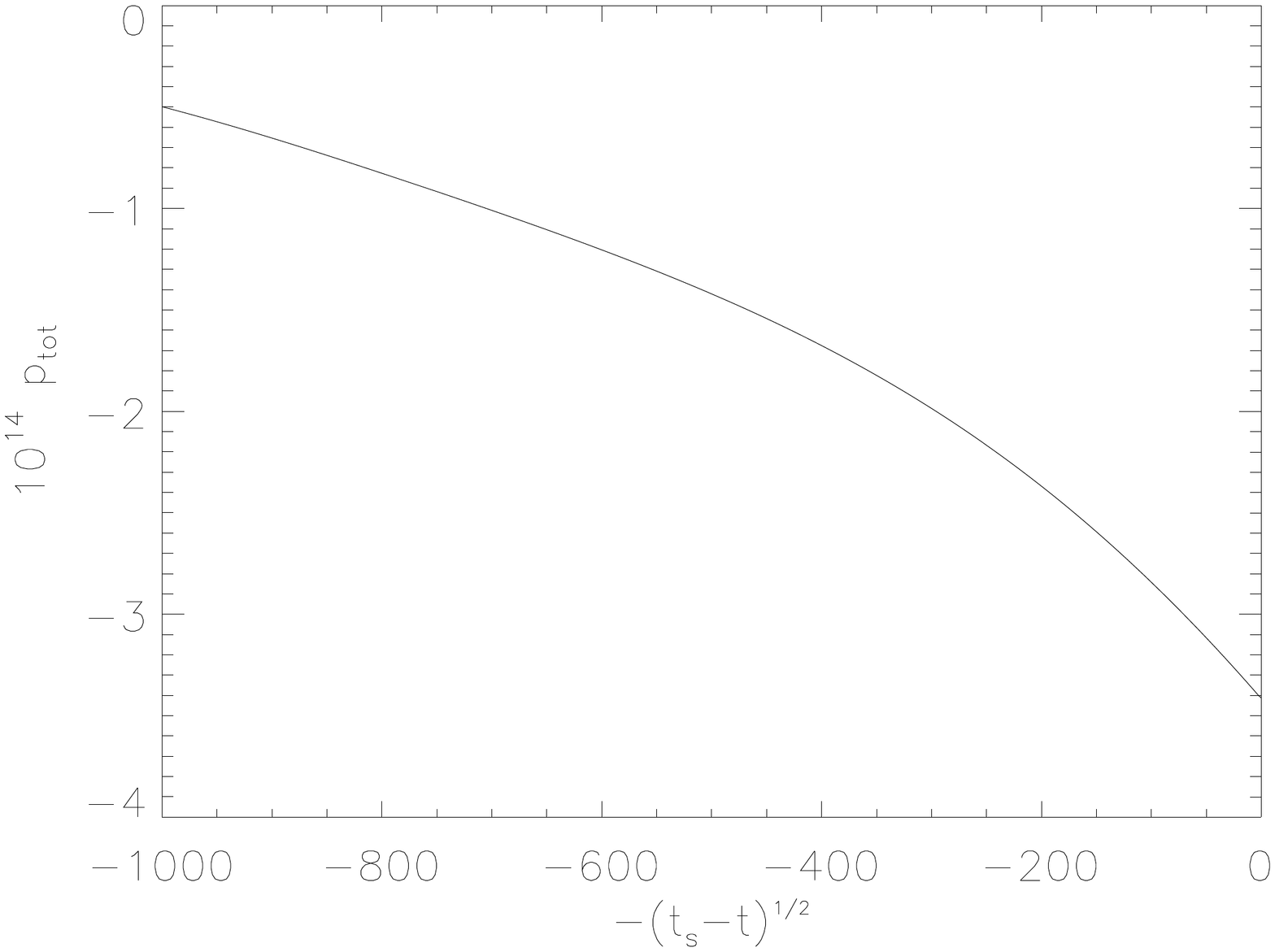}}
\caption{Classical type III singularity with $\alpha=10^9$ and $\beta=10^8$.  The plot on the left shows the energy density $\rho_{\rm tot}= \rho_{de} + \rho_{qe}$ for a solution to the semiclassical backreaction equation~\eqref{fq} for the case $B=\frc53$ and $A=3\times 10^9$.  The plot on the right shows the pressure $p_{\rm tot} = p_{de}+p_{qe}$ (with $p_{qe}$ obtained by
 substituting $\rho_{qe}$ into the conservation equation~\eqref{conserve}) for a solution to the semiclassical backreaction equations for the case $B=3$ and $A=5 \times 10^{28}$.  Note that in each case we are plotting the relevant behavior as a function of $-\sqrt{t_s-t}$, illustrating the expected linear behavior in terms of this quantity near the singularity.}
\label{Type_III_2and3_Fig}
\end{figure}

If $B>\frc32$, and $B \ne 2$, we find
that, near the singularity, solutions to~\eqref{fq} can again be expanded in powers of $t_s - t$, with the result
\be a = a_s \exp \left[ -c_1 \left(t_s-t\right) - c_2 \left(t_s-t\right)^2 - c_\gamma (t_s - t)^\gamma  - \cdots \right] \;,  \label{a-sol-2} \ee
with $c_1$ and $c_2$ arbitrary constants, and
\bes
\bea
\gamma &=& \frac{3B-4}{B-1} \; , \label{gamma3} \\
c_\gamma &=& \frac{1}{36\alpha \gamma (\gamma{-}1)(\gamma{-}2)}\left[3A(B{-}1)c_1^B\right]^\frac{-1}{B-1} \; , \label{cg2} \\
\rho_{de} &\approx& \left[3A(B{-}1)c_1(t_s-t)\right]^\frac{-1}{B-1} \; . \label{rq6}
\eea
\ees
For $\frc32<B<2$, we find $1<\gamma<2$, so that $\ddot a$ is the lowest derivative of $a$ that diverges (type II singularity), while for $2<B$, $2<\gamma<3$, so that $\dddot a$ is the lowest derivative of $a$ that diverges (type IV singularity). Figs.~\ref{Type_III_2and3_Fig}a and \ref{Type_III_2and3_Fig}b illustrate how $\rho_{\rm tot}$ and $p_{\rm tot}$ remain finite in these two cases.  Note that throughout this range, the leading term in \eqref{a-sol-2} is the $c_1$ term, which must be positive, since $H > 0$ at the singularity for $\alpha > 0$.  This again represents a softening of the classical type III singularity by quantum corrections, in this case by converting it to a type II or type IV singularity.

\subsection{Classical singularities of types II and IV}
\subsubsection{Analytic proofs regarding solutions to the semiclassical backreaction equations}
\label{sec:alpha-gt-0-type-II-IV-proofs}

If classically there is a type II or IV singularity and the singularity is at $a = a_s < \infty$, then $\rho_{de}(a_s)$ is finite.  For a classical type II singularity $p_{de}$,   $d\rho_{de}/da$ and all higher derivatives of $\rho_{de}$ diverge as $a \rightarrow a_s$.  For a classical type IV singularity, $p_{de}$ and $d \rho_{de}/da$ are finite, but all higher derivatives of $\rho_{de}$ diverge as $a \rightarrow a_s$.

In Sec.~\ref{sec:alpha-gt-0-type-III-proofs} it was pointed out that for a solution to the semiclassical backreaction equations~\eqref{fq-2} with a type II singularity, $d f/dy$ diverges
at the singularity but $f$ is finite.  If the solution has a type IV singularity then
$d^2 f/dy^2$ diverges at the singularity but $f$ and $df/dy$ are finite.

If classically there is a type II or type IV singularity, the first and third terms in the semiclassical backreaction equation~\eqref{fq-2} must be finite at the singularity since bounce solutions (for which $f \rightarrow 0$)
cannot occur for $\alpha >0$ and since $\rho_{de}$ is finite at the singularity.  Thus the only way for $d^2 f/dy^2$ to diverge at the singularity is for $f$ to diverge there.  In that case near the singularity~\eqref{fq-2} has the approximate form
\be \frac{d^2 f}{d y^2} = \frac{\beta}{12 \alpha} \frac{f}{y_s^2}   \;,  \label{fq-II-IV} \ee
with $y_s = a_s^3$.  The solutions to this equation are all finite at $y = y_s$.  Therefore it is not possible for $d^2 f/dy^2$ to diverge at the singularity if
there is a classical type II or IV singularity.

However, for a classical type II singularity $d\rho_{de}/da$ diverges as $a \rightarrow a_s$, so by taking one derivative of Eq.~\eqref{fq-2} it is easy to show that
$d^3 f/dy^3$ does diverge as $a \rightarrow a_s$.  This translates to the divergence of the fourth time derivative of $a$.  Similarly if there is a classical type IV singularity then $d^3 f/dy^3$ is finite but by taking two derivatives of Eq.~\eqref{fq-2}, it is easily seen that $d^4 f/dy^4$ diverges as $a \rightarrow a_s$.  This translates to a fifth time derivative of $a$. Thus in both cases there is a singularity but the singularity is softened to the point that we say it has been effectively removed.

\subsubsection{Specific analytic and numerical calculations}
\label{sec:num-alpha-gt-0-type-II-IV}

In Sec.~\ref{sec:classical-specific-type-II-IV} it was shown that classical type II singularities occur for the equation of state~\eqref{eq-state-de}
if $\rho_s > 0$ and $B < 0$ while there is a type IV singularity if $\rho_s > 0$ and $0 < B < \frc{1}{2}$.

  Solutions to~\eqref{fq} can be expanded about $t-t_s$ with the result
\be a = a_s \exp \left[ -c_1 \left(t_s-t\right) - c_2 \left(t_s-t\right)^2  - c_\gamma \left|t_s - t\right|^\gamma - \cdots \right] \;.  \label{a-sol-3} \ee
Here
\bes
\bea
\gamma &=&3+\frac1{1{-}B} \; ,\label{gamma4} \\
c_\gamma &=& \frac1{36\alpha\gamma(\gamma{-}1)(\gamma{-}2)} \left[3A(1{-}B)c_1^B\right]^\frac1{1-B} \;, \label{cg3} \\
\rho_{de} &\approx& \rho_s - \hbox{sgn}(t_s-t) \left[3A(1{-}B)c_1\left|t_s-t\right|\right]^\frac{1}{1-B} \; .\label{rq8}
\eea
\label{gamma-c-rho}
\ees
As predicted in Sec.~\ref{sec:alpha-gt-0-type-II-IV-proofs}, if there is classically a type II singularity, the lowest derivative of $a$ that diverges is $a^{(4)}$, and if there is classically a type IV singularity the lowest derivative that diverges is $a^{(5)}$.

\begin{figure}
\centering
\subfloat{\includegraphics[angle=90,width=3.1in,clip]{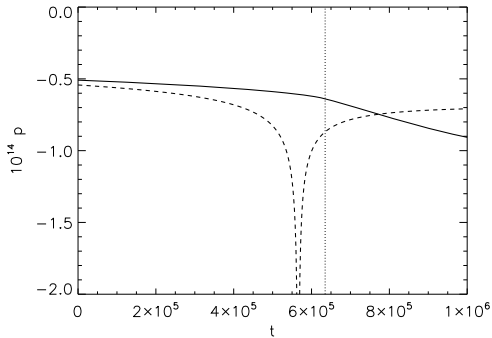}}\hfill
\caption{Classical type II singularity with $\alpha=10^9$, $\beta=10^8$ for $B=-1$, $A=10^{-30}$, and $\rho_s=5\times 10^{-15}$.  The dashed line denotes the pressure $p_{de}$ for a solution to the classical Einstein equations.  The solid line denotes the total pressure $p_{\rm tot} = p_{de}+p_{qe}$ (with $p_{qe}$ obtained by
 substituting $\rho_{qe}$ into the conservation equation~\eqref{conserve}) for a solution to the semiclassical backreaction equations.
 Note that for the classical solution the pressure diverges at the singularity.  The time at which the singularity occurs for the solution to~\eqref{fq} is denoted by the vertical dotted line.  At this time $p_{de} \rightarrow -\infty$ and $p_{qe} \rightarrow + \infty$ in such a way that $p_{\rm tot}$ remains finite at the singularity.  The continuous derivative of $p_{\rm tot}$ at the singularity implies that the singularity is weaker than a type IV singularity and thus has effectively been removed. }
\label{Type_II_Fig}
\end{figure}

The behavior for a type II model is illustrated in Fig.~\ref{Type_II_Fig}.    The plots for type IV models are very similar.

\section{Effects for $\alpha < 0$}\label{sec:PartB}

Though Starobinsky inflation requires $\alpha > 0$, for completeness we also want to consider the other possibilities.  We first note that when $\alpha<0$, solutions of~\eqref{fq} with $H=0$ do exist, and therefore it is possible for the universe to bounce, i.e.\ reach a maximum size and recontract, thus avoiding the singularity.  In some cases, avoiding the singularity is inevitable, while in others it is avoided only by a judicious choice of initial conditions.

\subsection{Classical little rips and big rip singularities}
\subsubsection{Proof that no little rip or big rip singularities occur}
\label{sec:alpha-lt-0-type-I-proof}

For classical big rip cosmologies with $p_{de} = w\rho_{de}$ when  $w<-1$ is a constant, an argument was given in~\cite{not} that solutions to~\eqref{fq} inevitably undergo bounces.
A calculation in~\cite{srivastava} and strong arguments in~\cite{comment,hae-1,haro} bolstered this conclusion.  Here we present a proof that formalizes and generalizes these arguments.  The generalization is to arbitrary equations of state $p_{de}(\rho_{de})$ which lead to classical little rips and classical big rip singularities.  As shown in Sec.~\ref{sec:alpha-lt-0-type-III-proof}, a similar proof works for arbitrary equations of state that lead to classical type III singularities.
The proof states that for $\alpha <0$, if $\rho_{de}$ diverges as $a \rightarrow \infty$, then all solutions to the semiclassical backreaction equations~\eqref{fq} must have $\dot{a} < \infty$ at all times, and that when the scale factor grows large enough there will be a bounce.  Thus it is impossible to have solutions with little rips, big rip singularities, or type III singularities.  Instead classical little rips and big rip singularities are always avoided if $\alpha <0$, and instead the universe undergoes a bounce.  As shown below, type III singularities are either avoided by a bounce or softened to type II or type IV singularities.

We begin by deriving an inequality which holds whenever $\alpha < 0$ and the energy density of the dark matter becomes large enough.
First we rewrite~\eqref{fq-2} in the form
\be\label{fq-4}
\alpha y \frac{d^2f}{dy^2} = \frac{\rho_{de}}{216}\left(\frac{y}{f}\right)^{5/3} + \frac{\beta}{12}\left(\frac{f}{y}\right) - \frac{1}{576\pi} \left(\frac{y}{f}\right)^{1/3} \; .
\ee
Using the arithmetic-geometric inequality, $\frc12(x+y) \ge \sqrt{xy}$, for the first two terms on the right in~\eqref{fq-4}, it is easy to see that the entire expression on the right is always non-negative if $\rho_{de} \ge \rho_0$, where
\be\label{rhomax}
\rho_0=\frac{1}{512\pi^2 \beta} \; .
\ee

The quantity $\rho_{de}(a)$ always diverges as $a \rightarrow \infty$ for equations of state which lead to classical little rips and big rip singularities.
In these cases there will be a constant $a_0$ such that $\rho_{de} \ge \rho_0$ for all $a \ge a_0$.  Then for $a > a_0$
we can put some positive lower bound $c_0$ on the right hand side of~\eqref{fq-4}.
 Keeping in mind that $\alpha < 0$, we therefore conclude that asymptotically for the classical little rip case or as the singularity is approached for the classical big rip case,
\be \frac{d^2f}{dy^2} \le -\frac{c_0}{|\alpha| y} \; . \label{proof-1} \ee
Integrating this twice, we find
\be f \le -\frac{c_0}{|\alpha|} y \log(y) +c_1 y +c_2 \; , \label{proof-2} \ee
where $c_1$ and $c_2$ are constants of integration.

Examination of~\eqref{proof-2} shows that $f$ and hence $\dot{a}$ vanishes for any solution to~\eqref{fq-2} which reaches a large enough value of $y$ and hence $a$.
Therefore if classically there is either a little rip or a big rip singularity the universe must undergo a bounce which prevents the limit $a \rightarrow \infty$
being achieved and therefore prevents either a little rip or a big rip from occurring. Since there is no singularity at the bounce, we say that the singularity is avoided.
It is possible to investigate what happens after the bounce, and this has been done for the case $p_{de} = w \rho_{de}$ with $w < -1$ in~\cite{hae-1, haro, hae-2}.
We do not include such an investigation here because our assumption that $\rho_{de}$ is the dominant form of the classical matter is not valid if the universe contracts to a small
enough size.  Furthermore, the properties that we assume for the dark matter only apply in the region of the final singularity.  The dark matter could have a different behavior
for smaller values of the scale factor.

\subsubsection{Specific analytic and numerical calculations}

For our specific models with equation of state given by~\eqref{eq-state-de}, we have numerically confirmed that classical little rips and big rip singularities are avoided by a bounce for a wide variety of parameters and initial conditions.

\subsection{Classical type III singularities}
\subsubsection{Proof that type III singularities are removed or softened}
\label{sec:alpha-lt-0-type-III-proof}

The proof given in~\ref{sec:alpha-lt-0-type-I-proof}, can easily be adapted to the case of equations of state $p_{de}(\rho_{de})$ which result in classical type III singularities.
First recall that in this case the scale factor has the finite value $a_s$ at the singularity while $\rho_{de}$ diverges in the limit $a \rightarrow a_s$.  Thus the conditions leading to the inequality~\eqref{proof-2} are satisfied as $a \rightarrow a_s$.
The inequality places an upper bound on $f = (a \dot a)^{3/2}$, so this implies $\dot a$ is also bounded.  Hence there are no corresponding solutions to the semiclassical backreaction equation~\eqref{fq} with type III singularities, and also clearly none with little rips or big rip singularities.  Instead, either the solutions in this
case must bounce when $a < a_s$, in which case the singularity is avoided, or the singularity at $a = a_s$ must be softened to a type II or type IV singularity.

To determine when a bounce must occur, note that the analysis leading to~\eqref{f-soln-4} still works for $\alpha <0$.  Thus if~\eqref{I2} diverges as the singularity at $a_s$ is approached, then $f \rightarrow - \infty$, since $\alpha < 0$, which is a contradiction since $f \ge 0$.
Thus for any model for the dark energy with a classical type III singularity for which~\eqref{I2} diverges, the corresponding solutions to the semiclassical backreaction equations~\eqref{fq} must bounce before the singularity is reached.

In all other cases there are two possibilities.  One is that a bounce can occur for $a < a_s$ in which case the singularity is avoided.  The other possibility is that $a \rightarrow a_s$
before a bounce occurs.  In this case the analysis in Sec.~\ref{sec:alpha-gt-0-type-III-proofs} still works and
the singularity is softened.  In particular, if~\eqref{I2} is finite but~\eqref{I1} diverges, then the singularity becomes a type II singularity, while if both are finite then the singularity becomes a type IV singularity.

\subsubsection{Specific analytic and numerical calculations}

For our specific equation of state~\eqref{eq-state-de}, recall that classical type III singularities occur
if $\rho_s=0$ and $B>1$.  The integrals $I_1$ and $I_2$ in~\eqref{I1} and~\eqref{I2} are the same regardless
of the values of $\alpha$ and $\beta$.  Thus that part of the analysis in Sec.~\ref{num-alpha-gt-0-type-III} remains the same.  There it was found that
both diverge if $1 < B \le \frc{3}{2}$.  What is different is that above we have shown that if $\alpha < 0$ then a bounce must occur before the singularity
is reached if both integrals diverge.

 For $B > \frc32$, the universe may reach a maximum size and bounce before reaching the singularity, it may reach the singularity while the universe is still expanding, or with finely tuned initial conditions, it may reach the singularity just as the universe stops expanding, so that $H=0$ at the singularity.  Singularities where $H=0$ were studied in~\cite{not,haro-2}, and we will not consider these further.
  If $H>0$ at the singularity  then the analysis in Sec.~\ref{num-alpha-gt-0-type-III} still holds and the behavior
   near the singularity can still be described by~(\ref{a-sol-2}) and (\ref{rq6}) with $c_1 >0$.  Thus for $B >\frc32$, if a bounce does not occur before $a \rightarrow a_s$, then the categorization of the singularities is the same as it is for $\alpha > 0$.

\subsection{Classical type II and IV singularities}

If $\alpha < 0$ then type II and IV singularities can, again, be avoided if a bounce occurs before the singularity is reached.  If they are not avoided then the
analysis in Sec.~\ref{sec:alpha-gt-0-type-II-IV-proofs} for $\alpha >0$ holds and the singularities are effectively removed.

For our specific equation of state~\eqref{eq-state-de} the analysis is similar to that in Sec.~\ref{sec:num-alpha-gt-0-type-II-IV}.
If the singularity is not avoided, then $H$ can be either positive or zero at the singularity.  The case where $H=0$ at the singularity was studied in~\cite{not,haro-2}, which again we will not consider.  If $H$ is positive then~\eqref{a-sol-3} and \eqref{gamma-c-rho} remain valid and the singularity is effectively removed.

\section{$\alpha = 0$}

The final case, $\alpha=0$, was studied in~\cite{haro-2}, where it was found that for a classical big rip model with $p_{de} = w \rho_{de}$, $\dot H$ diverges at a finite value of the scale factor resulting in a type II singularity.  In the general case, solving~\eqref{fq} for $H$ one finds
\be H^2=\frac{1\pm \sqrt{1-512 \pi^2 \beta \rho_{de}}}{96 \pi \beta} \; . \label{a0case} \ee
Note that choosing the minus sign gives the proper classical limit $H^2 = \frc83 \pi \rho_{de}$ as $\rho_{de} \rightarrow 0$.  For $\rho_{de} < \rho_0$, where $\rho_0$ is given by \eqref{rhomax}, $H^2$ is a smooth function of $\rho$, and hence if we have a type II or type IV singularity with $\rho_s < \rho_0$, there will be no change in the categorization of the resulting singularity at $\rho_{de}=\rho_s$. However,~(\ref{a0case}) becomes complex if $\rho_{de} > \rho_0$.
At $\rho_{de}=\rho_0$,~\eqref{a0case} implies that $H$ is still finite, but the derivative of the right hand side of~(\ref{a0case}) diverges, so that $\dot H$ diverges. Hence
for classical little rip cosmologies and classical big rip and type III singularities, the corresponding solutions to~\eqref{fq} all have type II singularities at $a = a_0$ with
$\rho_{de}(a_0) = \rho_0$.  This will also be the case for type II and type IV singularities with $\rho_s > \rho_0$.

\section{Conclusions}\label{sec:Conclusions}

We have considered the effects of both quantum fields and an $\alpha_0 R^2$ term in the gravitational Lagrangian on little rip models of the universe as well as future singularities of types I-IV.  Two methods have been used.  One is a background field approach where the effective stress-energy tensor for  a quantum field (and sometimes the $\alpha_0 R^2$ term) is computed in the background geometry.  The other involves finding solutions to the semiclassical backreaction equations.

Using a background field approach, the energy density for conformally coupled massive scalar fields has been computed for a particular fourth order adiabatic vacuum state in a particular spacetime with a type III singularity.  Similar calculations had previously been done for type I singularities.  In both cases the result is that the terms in the energy density that survive in the massless limit are the dominant ones near the singularity.  In that sense a massive scalar field becomes effectively conformally invariant near the singularity.  An argument was given that this behavior is likely to generalize to massive fields of spin $\frc{1}{2}$ and $1$, which are also conformally invariant in the massless limit, and it is likely to generalize to type II and IV singularities as well as little rip cosmologies.  This greatly increases the number of effectively conformally invariant fields near a future singularity.

A background field calculation of the energy density for conformally invariant fields and the effective energy density coming from an $\alpha_0 R^2$ term in the gravitational Lagrangian was computed for spatially flat Robertson-Walker spacetimes that are solutions to the classical Einstein equations with the dark matter as a source.  {As discussed in Sec.~II, some of the terms in the energy density for the quantum fields are the same as those for the  $\alpha_0 R^2$ contribution resulting in an effective coefficient for this term which we call $\alpha$.}    The energy density was expressed in terms of the energy density and pressure of the dark energy.  It was argued that quantum effects should be important near a future singularity well before the Planck scale is reached if $|\alpha|$ has a value comparable to that needed for Starobinsky inflation and/or there are enough quantum fields which are effectively conformally invariant near the singularity.

 We next investigated solutions to the semiclassical backreation equation~\eqref{fq} with an arbitrary number of conformally invariant fields together with the $\alpha_0 R^2$ term in the Lagrangian.  Because the nature of the dark energy is unknown, we made as few assumptions about it as possible.  The primary one was that it is a perfect fluid with equation of state $p_{de} = p_{de}(\rho_{de})$.  We then considered generic properties for the dark energy that would lead to solutions to the classical Einstein equations with little rips or final singularities of type I, II, III, or IV.  In each case we found general theorems which predict how solutions to~\eqref{fq} will behave for a given type of classical solution.

 If $\alpha > 0$ (the case in which Starobinsky inflation can occur) the singularity is never avoided because, as had been found previously, there are no bounce solutions for which the universe stops expanding and starts contracting.  We found that big rip singularities (also known as type I singularities) and little rip cosmologies always turned into big rip singularities, for which the scale factor diverges at a finite time in the future.  Classical type III singularities, in which the scale factor $a$ remains finite while its first time derivative $\dot a$ diverges, were modified in a model-dependent fashion.  These singularities remain type III singularities if the dark energy is such that the double integral in~\eqref{I2} diverges.  It this integral is finite but the single integral~\eqref{I1} diverges, then type III singularities become type II singularities.  Finally, if the single integral~\eqref{I1} is finite, then type III singularities become type IV singularities.
 Classical singularities of type II, in which $a$ and $\dot{a}$ are finite but higher derivatives diverge, are softened to the point that only $a^{(4)}$ and higher derivatives diverge.  Finally, classical singularities of type IV, in which only $a^{(3)}$ and higher derivatives diverge, are softened to the point that only $a^{(5)}$ and higher derivatives diverge.  Thus
 we say that both type II and IV singularities are effectively removed.

 If $\alpha < 0$ we found that classical big rip or little rip singularities are always avoided by a bounce.  Type III singularities for which the double integral in~\eqref{I2} diverges are also always avoided by a bounce.  All other singularities can be avoided by a bounce, depending on initial conditions.  If the singularity is not avoided, then classical type III singularities for which the double integral~\eqref{I2} is finite but the single integral~\eqref{I1} diverges are softened to type II singularities.  If both integrals are finite the singularity is softened to a type IV singularity.  Classical type II and IV singularities are effectively removed in the same way as occurs if $\alpha > 0$.

Finally, if $\alpha = 0$, we found that it is impossible for the density of the dark energy to become infinite as it does classically for little rips and type I and III singularities.  Instead a new type II singularity is created at a finite density $\rho_0$ given by~\eqref{rhomax}.  The same thing occurs for classical type II and IV singularities if the energy density $\rho_{de}$ at the classical singularity is larger than $\rho_0$.  If $\rho_{de}$ at the classical singularity is smaller than $\rho_0$, then type II and IV singularities remain type II and IV singularities.

To illustrate these results we used the three parameter equation of state of the dark energy given in~\eqref{eq-state-de}.  Different values and ranges of values of the parameters result in different types of final behaviors for the universe when the classical Einstein equations are solved with the dark matter as a source.  We solved  the full semiclassical backreaction equations~\eqref{fq} both analytically and, for certain values of the parameters, numerically.  We also solved the order reduced equations~\eqref{parker-simon-eq} numerically for certain values of the parameters. Not surprisingly, the order reduced method, which is perturbative in nature, produced effects that differed drastically from the full backreaction equations.  The reason is that the quantum effects become comparable to or even larger than the classical contributions, so it is not surprising that this approach breaks down. Some of our numerical results are shown in Figs.~1-5.

All of our backreaction results, including both the general proofs and the calculations for the specific equation of state~\eqref{eq-state-de}, are summarized in Table II.  The first column lists the types of final behaviors we considered for the solutions to the classical Einstein equations with the dark energy as a source.  The second through fourth columns list the behaviors of the corresponding solutions to the full semiclassical backreaction equations~\eqref{fq}.
Note that in situations labeled ``Bounce,'' the universe {\it must} bounce, independent of initial conditions.  In the rest of the cases with $\alpha < 0$, the singularity might be avoided by a bounce depending on initial conditions.  The singularities appearing in Table~I in this case apply {\it only} if the initial conditions do not lead to a bounce.
The last two columns give the parameter ranges for our specific model of the dark energy in~\eqref{eq-state-de} for which these different types of behaviors occur.

{\center
\begin{table}
\begin{tabular}{l c c c c c}
\hline\hline
~ & \multicolumn{3}{c}{Semiclassical modification} & ~ & ~ \\ \cline{2-4}
Classical classification & $\alpha > 0$ & $\alpha < 0$ & $\alpha =0$ & \multicolumn{2}{c}{~~~~~Specific eq. of state parameters} \\ \hline
Little rip & Big rip/type I & Bounce & Type II & ~~~~~~~~~~$\rho_s = 0$ & $B \le \frc12$ \\
Big rip/type I & Big rip/type I & Bounce & Type II & ~~~~~~~~~~$\rho_s = 0$ & $\frc12 < B \le 1$ \\
Type III &
$\left\{ \begin{matrix} \hbox{Type III} \\ \hbox{Type II} \\ \hbox{Type IV}\end{matrix} \right.$ &
$\begin{matrix} \hbox{Bounce} \\ \hbox{Type II} \\ \hbox{Type IV} \end{matrix}$ &
$\begin{matrix} \hbox{Type II} \\ \hbox{Type II} \\ \hbox{Type II} \end{matrix}$ &
~~~~~~~~~~$\begin{matrix} \rho_s=0 \\ \rho_s = 0 \\ \rho_s = 0 \end{matrix}$ &
$\begin{matrix} 1 < B \le \frc32 \\ \frc32 < B \le 2 \\ 2 < B \end{matrix}$\\
Type II & \multicolumn{2}{c}{Effectively removed} & Type II & ~~~~~~~~~~$\rho_s > 0$ & $B<0$ \\
Type IV & \multicolumn{2}{c}{Effectively removed} & Type II or IV & ~~~~~~~~~~$\rho_s > 0$ & $0 < B < \frc12$ \\ \hline\hline
\end{tabular}
  \caption{Modifications of various classical singularities due to conformally invariant fields and an $\alpha R^2$ term in the gravitational Lagrangian. The central columns give model-independent modifications for the three cases $\alpha > 0$, $\alpha < 0$, and $\alpha = 0$.  Note that when $ \alpha \ge 0$ reaching the singularity is inevitable; for $\alpha < 0$ it can always be avoided by a judicious choice of initial conditions, and for the cases marked ``Bounce'' the avoidance is guaranteed.  The final two columns give parameter ranges for our particular equation of state~\eqref{eq-state-de} that exhibit the various behaviors.}
  \label{restab}
\end{table}
}

\begin{acknowledgments}
PRA and BH would like to thank Jason Bates for helpful conversations and for sharing
with us a program he wrote to compute the energy density for massive minimally and
conformally coupled scalar fields in spatially flat Robertson-Walker spacetimes with
big rip singularities.  This work was supported in part by the National Science Foundation under Grants No. PHY-0856050, No. PHY-1308325, and No. PHY-1505875 to Wake Forest University.  Some of the numerical work was done using the WFU DEAC cluster; we thank the WFU Provost's Office and Information Systems Department for their generous support.
\end{acknowledgments}

\appendix

\section{Renormalized energy density for a conformally coupled massive scalar field}

In this appendix we derive the explicit form of the renormalized energy density for a massive conformally coupled scalar field that is displayed in~\eqref{rho-m}. Note that for this field the scalar curvature coupling constant is $\xi = \frc16$ .  In~\cite{a-e} the energy density for this field is written in the form
\bes \bea
  \la \rho_q \ra &=& \la \rho \ra_n + \la \rho \ra_{an}  \;, \\
  \la \rho \ra_n &=& \la \rho \ra_u - \la \rho \ra_{d} \;, \eea
  with\footnote{Note that there is a misprint in~(9a) of~\cite{a-e}.  The term in the third line which is proportional to $m^2$ should be multiplied by a factor of $(\xi - \frc16)$.}
\bea \la \rho \ra_u &=& \frac{1}{4 \pi^2 a^4} \int_0^\infty d k \, k^2 \left[ | \psi_k^{'}|^2 + (k^2 + m^2 a^2) |\psi_k|^2 \right] \;,  \label{rho-u} \\
  \la \rho \ra_d &=& \frac{1}{4 \pi^2 a^4} \int_0^\infty dk \, k^2 \left[ k + \frac{m^2 a^2}{2k} - \frac{m^4a^4}{8 k^3} \theta(k{-}\lambda)\right] \; , \label{rho-d} \\
  \la \rho \ra_{an} &=& \rho_a - \frac{m^4}{64 \pi^2} \left[ \frac{1}{2} + \log \left(\frac{m^2 a^2}{4 \lambda^2} \right) \right] \; ,  \label{rho-an}  \eea \ees
where $\theta(k{-}\lambda)$ is the Heaviside function, which introduces an arbitrary infrared cutoff at $\lambda$.
Note that $\rho_a$ is defined in~\eqref{rho-a}, and a prime denotes a derivative with respect to the conformal time $\eta$, which is defined by the relation
\be d \eta = \frac{d t}{a} \;. \label{conformal-time} \ee

To get $\la \rho_q \ra$ in the desired form, we use~\eqref{conformal-time} to convert to the proper time $t$ and use the definition in~\eqref{omega-def}.  Then
\be  \la \rho \ra_u = \frac{1}{4 \pi^2 a^2} \int_0^\infty d k \, k^2 \left[ | \dot{\psi}_k|^2 + \omega_k^2 |\psi_k|^2 \right] \;. \label{rho-u} \ee
Next, subtract and add back a term with the integrand proportional to $\omega_k$, so that
\be \la \rho_q \ra = \la \rho \ra_u - \frac{1}{4 \pi^2 a^3} \int_0^\infty dk \, k^2\omega_k   \; - \; \la \rho \ra_d +  \frac{1}{4 \pi^2 a^3} \int_0^\infty dk \, k^2\omega_k + \la \rho \ra_{an}  \;. \label{rhoq-app} \ee
Then
\bea - \la \rho \ra_d + \frac{1}{4 \pi^2 a^3} \int_0^\infty dk \, k^2\omega_k &=& \frac{1}{4 \pi^2 a^4} \int_0^\infty dk \, \left[a k^2 \omega_k - k^3 - \frac{m^2 a^2}{2} k + \frac{m^4 a^4}{8 k} \theta(k {-} \lambda) \right] \nonumber \\
  &=&  \frac{m^4}{64 \pi^2} \left[ \frac{1}{2} + \log \left(\frac{m^2 a^2}{4 \lambda^2} \right) \right]  \;. \label{rhod-subt} \eea
Substituting~\eqref{rho-an},~\eqref{rho-u}, and~\eqref{rhod-subt} into~\eqref{rhoq-app} gives the result in~\eqref{rho-m}.

\section{Adiabatic matching}

In this appendix we review the method of adiabatic matching~\cite{b-d-book} and show how it was used to specify the states for the massive conformally coupled scalar field that
were used in our numerical calculations.

It is easiest to begin by changing variables to $f_k = a^{1/2} \psi_k$.  Substituting into~\eqref{psi-eq} gives
\be \ddot{f}_k  + \left(\omega_k^2 - \frac{\ddot{a}}{2 a} + \frac{\dot{a}^2}{4 a^2}\right) f_k = 0  \;. \label{f-eq} \ee
Then a WKB approximation can be obtained with the variable transformation
\be f_k = \frac{1}{\sqrt{2 W_k(t)}} \, \exp \left( -i \int_0^t d \bar{t} \,  W_k(\bar{t}) \right) \;. \ee
Substituting into~\eqref{f-eq} one finds
\be W^2 =  \omega_k^2 - \frac{\ddot{a}}{2 a} + \frac{\dot{a}^2}{4 a^2} - \frac{1}{2} \left( \frac{\ddot{W}}{W} - \frac{3}{2} \frac{\dot{W}^2}{W^2} \right)  \;. \label{W-eq} \ee
Iterating with the zeroth order term given by $W_k^{(0)} = \omega_k$ gives for the second order term
\be W^{(2)} = \left[ \omega_k^2  -\frac{\ddot{a}}{2 a} + \frac{ \dot{a}^2}{4 a^2} - \frac{ \ddot{W}^{(0)}}{2 W^{(0)}} + \frac{3 (\dot{W}^{(0)})^2}{4 (W^{(0)})^2}
      \right]^{1/2} \;. \ee
Substituting into~\eqref{W-eq} gives
\be W^{(4)} = \left[ \omega_k^2  -\frac{\ddot{a}}{2 a} + \frac{ \dot{a}^2}{4 a^2} - \frac{ \ddot{W}^{(2)}}{2 W^{(2)}} + \frac{3 (\dot{W}^{(2)})^2}{4 (W^{(2)})^2}
      \right]^{1/2} \;. \ee

A fourth order approximation for $f$ evaluated at the time $t = 0$ is
\be f_k^{(4)} = \frac{1}{\sqrt{2 W^{(4)}_k(t)}} \;, \label{f-4} \ee
and one for $\dot{f}$ is
\be \dot{f}_k^{(4)} = -i\sqrt{\frac{W^{(4)}_k}{2}} - \frac{\dot{W}^{(2)}}{[2 W^{(2)}]^{3/2}} . \ee
By setting the exact mode function $f_k$ and its derivative $\dot{f}_k$ equal to these expressions at time $t=0$, one
fixes the state to be a fourth order adiabatic vacuum state and simultaneously provides starting values for the
numerical integration of the mode function.

\end{document}